\documentclass[preprint, authoryear]{elsarticle}

\usepackage{natbib}
\usepackage[utf8]{inputenc} 
\usepackage[T1]{fontenc}
\usepackage{lmodern}
\usepackage[fleqn]{mathtools}
\usepackage{booktabs}
\usepackage{rotating}
\usepackage{amssymb}
\usepackage{amsmath}
\usepackage{graphicx}
\usepackage{caption}
\usepackage{epstopdf}
\usepackage{stackrel}
\usepackage{soul}
\usepackage{color}
\usepackage{float}
\usepackage[official]{eurosym}
\usepackage{tabularx}
\usepackage{afterpage}
\usepackage{mathrsfs}
\usepackage{chngcntr}
\usepackage{caption}
\usepackage{subcaption}
\usepackage{verbatim}

\newcommand*\diff{\mathop{}\!\mathrm{d}}

\renewcommand*{\underline}{\ul}

\newdefinition{rmk}{Remark}
\newproof{pf}{Proof}
\newproof{pot}{Proof of Theorem \ref{thm2}}

\makeatletter
\def\ps@pprintTitle{%
	\let\@oddhead\@empty
	\let\@evenhead\@empty
	\def\@oddfoot{\centerline{\thepage}}%
	\let\@evenfoot\@oddfoot}
\makeatother

\begin{document}
\title{
	%On the statistical analysis \\ of a basket of cryptocurrencies 
	 %On the distribution function \\ of cryptocurrencies for risk assessment
	 On the Return Distributions of a Basket of Cryptocurrencies and Subsequent Implications
	%for inclusion in a market index.
	\\[12pt]
	%Clustering the cryptocurrency market using suitable (body- and tail-) return %distributions\\[12pt]
{\small(First Draft: May 14, 2021)}}
%\tnoteref{t1,t2}}
%\tnotetext[t1]{This document is a collaborative effort.}
%\tnotetext[t2]{The second title footnote which is a longer.}

\author[hhu]{Christoph J.\ Börner}
\ead{Christoph.Boerner@hhu.de}	
\author[hhu]{Ingo Hoffmann}
\ead{Ingo.Hoffmann@hhu.de}	
\author[hhu]{Jonas Krettek}
\ead{Jonas.Krettek@hhu.de}
\author[hhu]{Lars M.\ Kürzinger\corref{cor1}}
\ead{Lars.Kuerzinger@hhu.de}
\author[hhu]{Tim Schmitz}
\ead{Tim.Schmitz@hhu.de}

\cortext[cor1]{Corresponding author. Tel.: +49 211 81-11515; Fax.: +49 211 81-15316}
%\cortext[cor2]{Principal corresponding author}
%\fntext[fn1]{This is the specimen author footnote.}
%\fntext[fn2]{Another author footnote, but a little more longer.}
%\fntext[fn3]{Another author footnote, but a little more longer.}

\address[hhu]{Financial Services, Faculty of Business Administration and Economics, \\ Heinrich Heine University D\"usseldorf, 40225 D\"usseldorf,
	Germany}	

\begin{abstract}
		This paper evaluates and assesses the risk associated with capital allocation in cryptocurrencies (CCs). In this regard, we take a basket of 27 CCs and the CC index EWCI$^-$ into account. After considering a series of statistical tests we find the stable distribution (SDI) to be the most appropriate to model the body of CCs returns. However, as we find the SDI to possess less favorable properties in the tail area for high quantiles, the generalized Pareto distribution is adapted for a more precise risk assessment. We use a combination of both distributions to calculate the Value at Risk and the Conditional Value at Risk, indicating two subgroups of CCs with differing risk characteristics.

	%\end{description}

\end{abstract}

\begin{keyword}
%	\justifying
    Body-/Tail-Models \sep Cryptocurrencies \sep Index Construction \sep Market Segmentation \sep Statistical Tests\\[10pt]
	\noindent \textit{ORCID IDs:} 0000-0001-5722-3086 (Christoph J.~B\"orner), 0000-0001-7575-5537 (Ingo Hoffmann), 0000-0002-0978-6252 (Jonas Krettek), 0000-0001-5774-1983 (Lars M.~K\"urzinger), 0000-0001-9002-5129 (Tim Schmitz).\\[10pt]
	\noindent \textit{Acknowledgement:} We thank Coinmarketcap.com for generously providing the cryptocurrency time series data for our research. 
\end{keyword}
\maketitle
\newpage

\section{Introduction} \label{introduction}
Following the financial crisis of 2007 and a period of extreme uncertainty and volatility, trust in the financial system and its institutions, like central banks and their monetary policies, have been shattered \citep{KayaSoylu.2020, Bouri.2017b}.

Against this background, the market for cryptocurrencies (CCs) started to emerge in 2009 with the development of the Bitcoin by \cite{Nakamoto.2008}. This innovative peer-to-peer electronic cash system is not accountable to any subordinate institution, but is managed and controlled by its own community using the blockchain technology. Furthermore, the anonymity and security of transactions represent another noteworthy feature Bitcoin promises its users \citep{Kakinaka.2020}, resulting in increasing trading volumes and prices \citep{Corbet.2019}. 
This development raises questions for both investors and regulators alike regarding the CC market’s characteristics and risk profile, which need to be answered in order for CCs to become an investable asset class for a broad audience \citep{Gkillas.2018, Majoros.2018, Osterrieder.2017} and to provide guidance for risk management in general \citep{Kakinaka.2020}. In this regard, this study examines the question  which class of distribution functions models CC returns most accurately. We answer this question using a novel approach to separate a distribution's body from its tail proposed by \cite{Hoffmann.2021}. By doing so we are able determine the risk and statistical properties associated with CCs and provide valuable implications for portfoliomanagement and regulators alike.
 
Even though the technical properties of CCs are well understood, CCs’ behavior remains to be fully comprehended and analyzed. As of today, many economic studies have focused on Bitcoin and other prominent CCs such as Ethereum and Ripple since they dominate the CC market due to their high proportion of total market capitalization \citep{Glas.2019}. In this regard, \cite{Baur.2017} and \cite{Glas.2019} find Bitcoin and other CCs to be uncorrelated to traditional assets in times of financial distress. Additionally, \cite{Gkillas.2018d},  show CC’s behavior to differ from fiat currencies as well. Furthermore, various studies analyze certain characteristics of CCs like e.g.\ volatility \citep{Polasik.2015, Balcilar.2017}, diversification issues (see i.a.\ \cite{Briere.2015, Selgin.2015, Corbet.2018}) and safe haven properties \citep{Bouri.2017, Urquhart.2018}\footnote{For a more extensive literature review, see \cite{Corbet.2019}.}.

However, in order to evaluate the corresponding market risk and to completely understand the CC market as a whole, we follow other studies in analyzing return distributions of the CC market. As numerous studies observe non gaussian behavior and heavy tails in return distributions \citep{Osterrieder.2017, Gkillas.2018d, Gkillas.2018}, a distribution model accounting for these observed characteristics ought to be implemented. To account for these properties \cite{Majoros.2018} and \cite{Kakinaka.2020} utilize the class of stable distributions (SDIs) in their recent studies. However, \cite{Kakinaka.2020} observe the SDI to be unable to efficiently grasp the heavy tails of the analyzed return distributions in all scenarios in comparison to other possible distributions. Given this background, the generalized Pareto distribution (GPD), a statistical distribution which appears to model heavy tail properties more accurately, is used in further studies \citep{Gkillas.2018d, Gkillas.2018}. Following the approach presented in \cite{Hoffmann.2021}, we therefore attempt to use a combination of both described distributions. Analytically discovering the beginning of the tail of the analyzed return distributions, enables us to split the given data into a body and a tail for each of which we implement a different distribution. For the first sample, containing the tail of potential losses, we implement the GPD, since literature and our conducted tests show its goodness of fit for estimating tail values. For the remaining body of our data we apply the SDI as its fit outperforms other possible distributions. 

Our study aims to add to the existing literature by implementing a novel approach, which tries to reach a higher quality of modelling return distributions observed in the CC market. Furthermore, as of today most studies are merely concerned with analyzing characteristics of Bitcoin or the most prominent CCs like i.a.\ Ethereum, Ripple and Litecoin (\citep{Baur.2017, Bouri.2017, Osterrieder.2017, Gkillas.2018d, Gkillas.2018, Majoros.2018, Kakinaka.2020}. Therefore, most studies do not consider the entirety of the CC market, which, as \cite{Glas.2019} points out, might lead to potential bias. Hence, we join \cite{Glas.2019} and \cite{ElBahrawy.2018}, in an attempt to give a broader overview of the CC market. By doing so we tackle two existing gaps in literature found by \cite{Corbet.2019} in their extensive literature review.  Namely, we extend the number and size of analyzed data in an attempt to analyze CCs as an asset class. Furthermore, our reserach does provide practical relevance in form of an improved risk assessment. By analytically separating a distribution's body from its tail and implementing different distributions, we are able to estimate risk measures in form of the Value at Risk and the Conditional Value at Risk more precisely. Both of which do yield an importance for regulators and investors alike.

\begin{comment}The added value of this study compared to the existing literature lies in the considered market breadth, which is well covered by 27 crypto currencies. Recently completed research focuses on high frequency intraday data \citep{Majoros.2018, Kakinaka.2020}. In contrast to this, we extend the examined time scale in the present study to the weekly period. The latter will become interesting in future for institutional investors in terms of investment goals and risk assessments if they allocate a significant share in the investment portfolio to CCs. The main development of this study is the identification of stable distribution as a suitable model for the distribution function of weekly CCs returns on the basis of a market-wide analysis. \end{comment}
\begin{comment}The identification of a suitable distribution model is essential as a basis for a well-founded portfolio theory. The present study provides a concrete answer to this question.  Especially in the area of strategic asset allocation, i.e.\ if a large number of different asset classes are considered in the medium to long-term time range, the identified stable distribution will play a fundamental role. 
We also found the modeling quality limited in the tail and compared the modeling quality with the GPD as a tail model. Just like the work on high-frequency data, our results for the present data set of weekly returns also show that in practice the tail model mentioned above should be used instead of the body model for risk assessment in the case of high quantiles. \end{comment}

The remainder of this paper is structured as follows: In Sec.\ \ref{dataCC} we present and describe our data used for our following analysis. In Sec.\ \ref{distCC} we perform a series of statistical analyses and tests which lead us to the family of SDIs as the best model choice for the body of the CCs returns distribution.
Based on these findings, Sec.\ \ref{tailrisks} is concerned with the assessment of the tail risk inherent in an investment in a single CC or the basket aggregated in the EWCI$^-$. We compare both methods of the risk assessment at high quantiles. First we use the body model and second we adapt the GPD as tail model for risk assessment in terms of the Value at Risk and the Conditional Value at Risk. The last section summarizes our most import results and gives an overview of further research topics.

\section{Data} \label{dataCC}
As a foundation of our analysis, we follow various studies, by extracting CCs’ daily prices from the website coinmarketcap.com (i.a.\ \citep{Fry.2016, Hayes.2017, Brauneis.2018, Caporale.2018, Gandal.2018, Glas.2019}. For an observation period from 2014-01-01 to 2019-06-01, we take $N = 66 $ CCs from the Coinmarketcap Market Cap Ranking at the reference date of 2014-01-01 into consideration, cf.\ Tab.\ \ref{tab:DatasetTestAssetNutshell}.

In order to depict the CC market as a whole, we aim to include as many CCs in our analysis as possible. However, as data gaps appear in the time series of most CCs considered, we exclude all CCs with five or more consecutive missing observations. By utilizing the Last Observation Carried Forward (LOCF) approach, as previously done in \cite{Schmitz.2020} and \cite{Trimborn.2020}, we are able to include all CCs with smaller data gaps.

After taking these considerations into account, 27 CCs remain in our data set, all of which are depicted in bold letters in cf.\ Tab.\ \ref{tab:DatasetTestAssetNutshell}.
In a next step, we convert the CC prices denoted in USD to EUR prices, using the daily USD-EUR exchange rates retrieved from Thomson Reuters Eikon. For the purpose of preventing potential weekday biases, the resulting (daily) observations are converted to weekly observations in a following step. In a final step, we calculate logarithmic returns using these weekly CC EUR prices for each CC, which we will refer to as returns in the remainder of this paper.

\begin{table}[H]
	\footnotesize \noindent
	\begin{tabularx}{\textwidth}{  l  X   l  X l  X}
		\toprule
		CC & ID &  CC & ID & CC & ID\\
		\midrule
		{\bf Anoncoin} 		& ANC &
		Argentum 				& ARG &
		AsicCoin 				& ASC \\
		BBQCoin 				& BQC &
		BetaCoin 				& BET &
		{\bf BitBar} 			& BTB \\
		{\bf Bitcoin} 			& BTC &
		BitShares PTS 			& PTS & 
		Bullion 				& CBX \\ 
		ByteCoin 				& BTE &
		{\bf CasinoCoin} 		& CSC &
		CatCoin 				& CAT \\
		Copperlark 			& CLR &
		CraftCoin 				& CRC &
		Datacoin 				& DTC \\
		{\bf Deutsche e-Mark} 	& DEM &
		Devcoin 				& DVC & 
		{\bf Diamond} 			& DMD \\
		{\bf Digitalcoin} 		& DGC &
		{\bf Dogecoin} 		& DOGE& 
		Earthcoin 				& EAC \\ 
		Elacoin 				& ELC &
		EZCoin 				& EZC &
		FastCoin 				& FST \\ 
		{\bf Feathercoin} 		& FTC &
		Fedoracoin 			& TIPS&
		{\bf FLO} 				& FLO \\
		Franko 				& FRK &
		{\bf Freicoin} 		& FRC & 	 
		Globalcoin 			& GLC \\
		{\bf GoldCoin} 		& GLC & 
		GrandCoin 				& GDC &
		HoboNickels 			& HBN \\
		I0Coin 				& I0C &
		{\bf Infinitecoin} 	& IFC &
		Ixcoin 				& IXC \\
		Joulecoin 				& XJO &
		Junkcoin 				& JKC &
		{\bf Litecoin}  		& LTC \\
		LottoCoin 				& LOT &
		Luckycoin 				& LKY &
		{\bf Megacoin} 		& MEC \\   
		MemoryCoin 			& MMC & 
		MinCoin 				& MNC &
		{\bf Namecoin} 		& NMC \\
		NetCoin 				& NET &
		Noirbits 				& NRB &
		{\bf Novacoin} 	    & NVC \\
		{\bf Nxt} 				& NXT &
		{\bf Omni} 			& OMNI&
		Orbitcoin 				& ORB \\
		{\bf Peercoin} 		& PPC &
		Philosopher Stones 	& PHS &
		Phoenixcoin 			& PXC \\
		{\bf Primecoin} 		& XPM &
		{\bf Quark} 			& QRK &
		{\bf Ripple} 			& XRP \\
		SexCoin 				& SXC &
		Spots 					& SPT &
		StableCoin 			& SBC \\
		{\bf TagCoin} 			& TAG &
		{\bf Terracoin} 		& TRC &
		Tickets 				& TIX \\
		TigerCoin 				& TGC &
		{\bf WorldCoin} 		& WDC &
		{\bf Zetacoin} 		& ZET  \\
		
		\bottomrule
	\end{tabularx}
	\caption{: Considered CCs and depicted data set (bold type), data source: CoinMarketCap.}
	\label{tab:DatasetTestAssetNutshell}
\end{table}

For an aggregated perspective on CCs, we use the above-mentioned weekly CC data to calculate an Equally-Weighted CC Index (EWCI), as it is similarly done in \cite{Schmitz.2020}. Whereas, as we exclude more CCs in comparison to \cite{Schmitz.2020}, we will call this index \textit{EWCI$^-$} for a more precise distinction.

\section{The return distribution of cryptocurrencies} \label{distCC}
For an initial classification of CCs simple statistical key figures from the standard repertoire of empirical statistics are used below. The description and evaluation of additional statistical properties of CCs, for example Value at Risk or lower-partial moments, are carried out using a suitable distribution function. 
\begin{comment}These distribution functions cannot be derived from a theoretical concept and are therefore not readily available. \end{comment} 
Our results show that the family of SDIs is a suitable model for the examined returns of CCs. Hence, this family of distributions is used in Sec.\ \ref{tailrisks} for a more in-depth analysis of the statistical properties of CCs.

\subsection{Determination of basic statistical key figures of the cryptocurrencies}
\label{KeyFigures}
Standard procedures lead us to estimates of the set of basic statistical key figures: mean $\hat\mu$, variance $\hat\sigma^2$ and the bandwidth (Tab.\ \ref{TabelleBasicStatistik}.).
\label{statkennzahlen}
\begin{table}[H]
	\footnotesize \noindent
	\begin{center}
	\begin{tabularx}{0.9\textwidth}{l  r  r  r  r  r r r r} \toprule
	Crypto								&	
	Mean								&	
	Variance							&	
	\multicolumn{2}{l}{Bandwidth}		& 
	\multicolumn{2}{l}{HDS-Test}		&	
	\multicolumn{2}{l}{SIG-Test}		\\
	ID 									&	
	${\hat \mu }$						&	
	${\hat \sigma^2 }$ 					&	
	{\tiny min} 						&	
	{\tiny max} 					   		&	
	H0							  		&	
	$p$-Value							&	
	H0									&	
	$p$-Value							\\ 	\midrule
	EWCI$^-$&	0.3	&	1.6	&	-40.8	&	37.7	&	0	&	79.2	&	0	&	10.8	\\
	ANC  &	-1.0	&	14.3	&	-241.4	&	162.9	&	0	&	99.0	&	0	&	17.1	\\
	BTB  &	-0.7	&	11.7	&	-201.2	&	150.4	&	0	&	97.4	&	0	&	37.2	\\
	BTC  &	0.9	&	1.0	&	-30.4	&	42.2	&	0	&	99.8	&	0	&	59.2	\\
	CSC  &	-1.2	&	30.5	&	-697.9	&	169.0	&	0	&	92.2	&	0	&	51.2	\\
	DEM  &	-1.4	&	9.3	&	-100.8	&	145.8	&	0	&	99.6	&	0	&	85.8	\\
	DMD  &	0.0	&	4.1	&	-84.7	&	102.1	&	0	&	99.0	&	0	&	43.9	\\
	DGC  &	-1.7	&	9.0	&	-217.4	&	116.2	&	0	&	96.6	&	0	&	31.1	\\
	DOGE &	0.9	&	3.7	&	-60.8	&	144.9	&	0	&	97.6	&	1	&	0.1	\\
	FTC  &	-0.9	&	7.6	&	-144.7	&	171.7	&	0	&	83.2	&	0	&	10.8	\\
	FLO  &	0.7	&	6.8	&	-67.9	&	162.2	&	0	&	84.8	&	0	&	43.9	\\
	FRC  &	-0.5	&	21.3	&	-332.1	&	338.9	&	0	&	100.0	&	0	&	95.3	\\
	GLC  &	0.2	&	5.6	&	-74.1	&	91.0	&	0	&	100.0	&	0	&	51.2	\\
	IFC  &	-0.6	&	10.6	&	-126.4	&	296.1	&	0	&	50.0	&	0	&	51.2	\\
	LTC  &	0.6	&	2.1	&	-34.2	&	87.5	&	0	&	99.0	&	0	&	21.1	\\
	MEC  &	-1.6	&	5.6	&	-112.9	&	133.1	&	0	&	96.6	&	0	&	85.8	\\
	NMC  &	-0.9	&	3.0	&	-110.2	&	82.4	&	0	&	100.0	&	0	&	25.8	\\
	NVC  &	-1.0	&	5.5	&	-235.7	&	128.2	&	0	&	99.6	&	0	&	8.4	\\
	NXT  &	-0.1	&	4.2	&	-83.9	&	106.5	&	0	&	93.8	&	0	&	8.4	\\
	OMNI &	-1.4	&	5.4	&	-73.1	&	116.9	&	0	&	82.6	&	0	&	43.9	\\
	PPC  &	-0.9	&	2.7	&	-60.2	&	73.8	&	0	&	86.8	&	0	&	21.1	\\
	XPM  &	-1.0	&	4.1	&	-67.7	&	117.7	&	0	&	91.2	&	1	&	4.9	\\
	ORK  &	-1.1	&	7.2	&	-94.1	&	137.9	&	0	&	96.8	&	0	&	37.2	\\
	XRP  &	1.1	&	3.8	&	-72.9	&	109.7	&	0	&	100.0	&	1	&	0.0	\\
	TAG  &	-1.1	&	5.3	&	-63.6	&	136.3	&	0	&	100.0	&	0	&	59.2	\\
	TRC  &	-1.0	&	6.7	&	-72.7	&	162.6	&	0	&	94.6	&	0	&	17.1	\\
	WDC  &	-1.6	&	6.8	&	-121.7	&	110.3	&	0	&	99.8	&	0	&	51.2	\\
	ZET  &	-1.0	&	6.6	&	-97.7	&	131.4	&	0	&	94.8	&	0	&	95.3	\\ \bottomrule
	\end{tabularx}
	\end{center}
	\caption{: Basic statistical key figures and tests on unimodality and symmetry. Units in per\-cent and boolean, see text.}
	\label{TabelleBasicStatistik}
\end{table}
It can be seen that the returns scatter strongly around a centre close to zero. While the variance and thus the standard deviation indicate leptokurtic behavior and therefore a concentration of returns, the sometimes considerable bandwidth indicates a strong blur of returns over a wide measurement range. This leads to the preliminary conclusion that the returns of CCs in the middle value range follow a concentrated distribution that has pronounced fat tails in the outer areas.
In particular, the large variance ($\sim 7 \%$) and the bandwidth ($\sim 300 \%$) of CCs clearly show the completely different character of CCs compared to traditional asset classes. Comparable values on a weekly basis for the traditional asset classes (stocks, bonds, real estate, etc.) fall in the range of $\sim 0.02\%$ (variance) or $\sim 0.1\%$ (bandwidth) on average.
Hence, in comparison  it can be seen that there is a considerable risk associated with CCs.

Additionally, in this step we use the Hartigan dip test \citep{Hartigan.1985} to test the null hypothesis H0 that the empirical distribution is unimodal and symmetric. So for each data set the Hartigans Dip Statistics (HDS) are calculated and evaluated. 
To test for symmetry, the simple sign test, cf.\ e.g.\ \cite{Gibbons.2011}, is carried out for each data set.
The last four columns in Tab.\ \ref{TabelleBasicStatistik} show the results of the both tests.\footnote{Note that for all tests performed in this paper, the following applies: The boolean "0" indicates that the null hypotheses cannot be rejected on a 5\%-level and alternatively the boolean "1" indicates a rejection.}

The results and especially the high $p$-values of the HDS-Test strongly suggest all data sets to obey a unimodal distribution. The CC Infinitecoin (IFC) shows the lowest $p$-value. This indicates that the empirical distribution could be multimodal. In fact, the histogram of returns for the IFC suggests a multi-peak nature. There are no indications of a fundamental structural break, so this tends to be more of a random nature and is due to insufficient statistics (cf.\ theorem of \citet{Glivenko.1933, Cantelli.1933}), which might occur with short samples in particular. When adjusting a unimodal distribution function later in Sec.\ \ref{sdiCC}, we expect a lower quality of the distribution model for the returns of this currency.

Apart from three exceptions, a clear result can also be seen in the SIG-Test for symmetry of the empirical distribution of returns. For the vast majority of CCs, the assumption of a symmetrical distribution of returns at a moderately high level of significance cannot be rejected.
While the assumption of a symmetrical distribution of the returns is narrowly rejected for the CC Primecoin (XPM), the rejection for the CCs Dogecoin (DOGE) and Ripple (XRP) is almost clear. We therefore assume that the distributions are slightly skewed.\begin{comment}
When fitting symmetrical distribution functions, the quality will be worse than with distribution functions that also allow slight skewness. This indicative result will be confirmed for the three currencies in the further analyzes in Sec.\ \ref{sdiCC}, cf.\ Tab.\ \ref{Tabelle_ADTest}. \\
\end{comment}

\begin{comment}
* This ties in with results earlier found by {\color{red} A, B, C, ... (Lit-Zitate)}\\
\end{comment}

Going forward, we assume the returns of the CCs to be concentrated around zero and the empirical distribution to have a fat tail due to the large bandwidth. Furthermore, we expect the empirical distribution to have a unimodal and essentially symmetrical shape. Anyway, it cannot be ruled out that the data sets in question may be (slightly) skewed. We will take this fact into account when selecting and adapting a suitable distribution function in Section \ref{sdiCC}.

\subsection{Statistical tests to further reduce the variety of possible distributions} \label{stattests}
A large number of mathematical models are available for the statistical description of CC returns. On the basis of some characteristic features of the data set, the family of models can be narrowed down and a suitable family of functions for representing the distribution can be deduced. In the following a series of statistical tests are carried out in order to infer possible function families for the description of our data sets. The same tests are also carried out for the EWCI$^-$ defined in Sec.\ \ref{dataCC}. In total, $N = 28$ time series are considered in the tests described below.

Overall, the statistical tests in Sec.\ \ref{KeyFigures} and the following are used to examine whether the combined hypothesis that the data sets have a unimodal, symmetrical and stationary distribution must be rejected. Furthermore, it is checked whether the hypothesis of an independent, identical distribution (IID) of the individual returns must be rejected, which is an important property required to specify a distribution function.

The augmented Dickey-Fuller test \citep{Dickey.1979, Wooldridge.2020} is used to test a possible rejection of the stationarity hypothesis. Finally, the ARCH test according to Engle \citep{Engle.1982, Engle.2002} is used to check whether the hypothesis of homoscedasticity of the innovation process $\epsilon_t$ for the individual CCs and the EWCI$^-$ must be rejected.

\subsubsection*{Augmented Dickey-Fuller Test}
The augmented Dickey-Fuller test (ADF) is performed considering the autoregressive model for the CC return time series, $y_t$, of each CC\footnote{ Note, the returns $r_t$ are calculated in this notation according to $r_t = y_t - y_{t-1}$.}: 
\begin{flalign}\label{ADFModell}
y_t & = c +  \delta t + \varphi y_{t-1} + \beta_1\Delta y_{t-1} 
		+ \beta_2\Delta y_{t-2} + \ldots + \beta_p\Delta y_{t-p}
		+ \epsilon_t.
\end{flalign}
With drift coefficient $c$, deterministic trend coefficient $\delta$,  AR(1) coefficient $\varphi$ and coefficients $\beta_i$ for the lag terms $i = 1, \ldots, p$ up to the order $p = 10$.  In Eq.\ (\ref{ADFModell}) $\epsilon_t$ denotes the innovation process. The aim of the test is to check the hypothesis of trend stationarity, i.e.\ $ \delta = 0$ is the null hypothesis, in the tables denoted by H0, see e.g.\ \citet{Wooldridge.2020}.

\begin{figure}[htbp]
	\centering
	\captionsetup{labelfont = bf, labelsep = none}
	\includegraphics[width=0.85\textwidth]{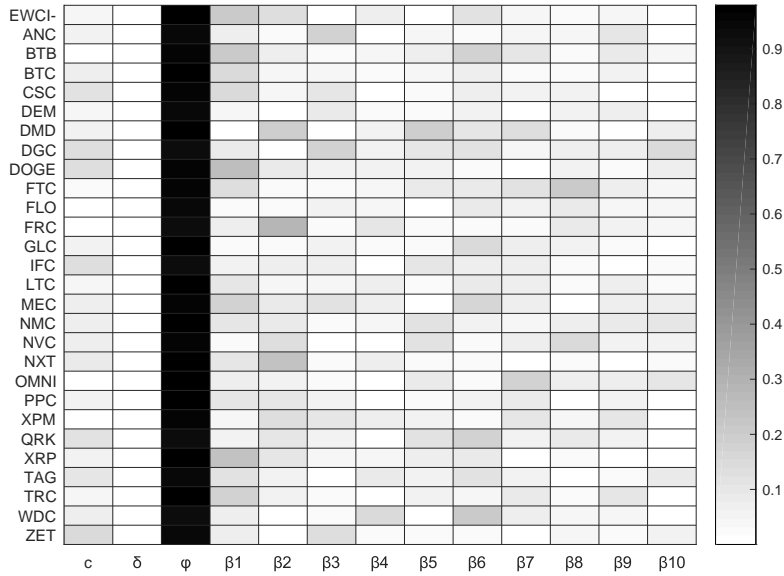}
	%\begin{quote}
	\caption[Heatmap]{\label{Fig3HeadmapLags} The Heatmap reflects the structure of the autoregressive model with drift coefficient $c$, deterministic trend coefficient $\delta$, AR(1) coefficient $\varphi$ and coefficients $\beta_i$ for the lag terms up to ten time shifts for each CC.}
	%\end{quote}
\end{figure}

The Heatmap in Fig.\ \ref{Fig3HeadmapLags} visualizes the structure of the fitted autoregressive model Eq.\ (\ref{ADFModell}). 

We find the deterministic trend coefficient $\delta$ to be comparable to zero in all CCs considered. The results in Tab.\ \ref{TabelleStattest} in the first four columns provide a deeper insight. 
For the vast majority of CCs, the $p$-values are comfortably high, so that a rejection of the null hypothesis of trend stationarity is not indicated here.
But as can be seen in the table the ADF-Test rejects the null hypothesis for the CCs FLO (FLO), Quark (QRK) and Worldcoin (WDC) on a 5\% confidence level.  
At this point we take a closer look at the corresponding trend coefficients: 
$\delta_{\text{FLO}} = 1.6$e-03, $\delta_{\text{QRK}} = 1.5 $e-04 and $\delta_{\text{WDC}} = 1.0$e-04. Since all the coefficients $\delta$ are close to zero, the influence of a possible trend is likely to be of significantly less importance.
Therefore, in the following, we assume trend stationarity ($\delta = 0$) for the time series of CCs.

The third column in the Heatmap in Fig.\ \ref{Fig3HeadmapLags} illustrates the value of $\varphi$. We find the parameter $\varphi$ to be greater than 0.9 for all CCs and, for the most part, clearly close to 1. The latter is an important condition to be fulfilled for the assumption of a random walk ($\varphi = 1$).

The augmented Dickey-Fuller test was performed for lags $p$ up to order ten. More complicated dynamics with serial correlation make themselves noticeable in the analysis by the fact that the coefficients of the terms corresponding to the lags are clearly different from zero. 
We find the absolute values of the coefficients (that means $\left| \beta_i\right|$) to be close to zero. In fact, the majority of the absolute coefficients $\left| \beta_i\right|$ are clearly smaller than 0.2 as an upper limit\footnote{By similar argumentation as in \citet[Chapter 11 therein]{Wooldridge.2020} a repeated substitution causes the effectiveness of the corresponding terms to fall below the 5\% mark in the next time step and thus become largely insignificant.  } 
and a basic model $y_t = c + y_{t-1} + \epsilon_t$ can be assumed (in Tab.\ \ref{TabelleStattest} third column denoted with "G"). Only a few single coefficients exceed the value 0.2 by a small margin and in these cases a model with a slightly influencing lag structure $y_t = c + y_{t-1} + \bar\beta\times(\text{Lag-Structure}) + \epsilon_t$ with an average coefficient $\bar\beta = 0.03$ could be guessed at. By marking the corresponding CCs with "L", Tab.~\ref{TabelleStattest} shows for which CCs this is the case.

 \begin{comment} i.e.\ very small $\bar\beta$ \end{comment}
 Due to the observed insignificance of the lag stucture, we assume a basic model "G" in these cases as well and expect statistical inaccuracies to distort the result due to the limited length of the time series. We therefore postulate a possible serial correlation in the data sets to be of minor importance.
Thus, the detailed analysis of the results of the augmented Dickey-Fuller test suggests 
that the model $r_t = c + \epsilon_t$ cannot not be rejected for the returns $r_t$. Here, $c$ denotes the individual time-constant drift term for each CC and, as above, $\epsilon_t$ denotes the innovation process.

\begin{comment}
We proceed from this possible basic model for the logarithmic returns $r_t$ in the following. The results of the augmented Dickey-Fuller test show however, that more complex models are possible in practice, which can be identified and validated more clearly if larger data sets were available. 
In practice, the possibility of a pronounced lag structure must be observed, checked and taken into account separately in individual cases. 
A thorough and critical review of the possibility of a distinct lag structure is essential and must be assessed and scrutinized, especially in risk assessmen,which is likely to get increasingly requested by regulatory auditing authorities in the future.
\end{comment}

\subsubsection*{Test on autoregressive conditional heteroscedasticity}

\begin{comment}
It is possible that the innovation process $\epsilon_t$ is a function of the actual sizes of the previous time periods' error terms. The analysis of the time series of logarithmic returns would then lead to the conclusion that there is an autoregressive conditional heteroscedasticity (ARCH). The ARCH-Test due to \citet{Engle.1982} is used to investigate this property. In this context, the following model of the innovation process is examined:
\begin{flalign}\label{ARCHModell1}
\epsilon_t & = \sigma_t z_t.
\end{flalign}
The random variable $z_t$ is a strong white noise process and the series $\sigma_t$ is modeled by:
\begin{flalign}\label{ARCHModell2}
\sigma_t^2 & = \alpha_0 + \alpha_1\epsilon_{t-1}^2 + \ldots + \alpha_q\epsilon_{t-q}^2.
\end{flalign}
With coefficients $\alpha_0 > 0$ and $\alpha_i \geq 0$ for $i>0$. If $\alpha_i = 0$ for $i>0$ there is no dynamics in the variance equation. Hence, we have a constant variance $\sigma^2 \coloneqq \sigma_t^2 = \alpha_0$, see e.g.\ \citet{Wooldridge.2020}.
The null hypothesis is that we have $\alpha_i =0$ for all time shifts $i = 1, \ldots, q$. The alternative hypothesis is that at least one of the estimated $\alpha_i$
coefficients in Eq.\ (\ref{ARCHModell2}) must be signifiant.
\end{comment}

The ARCH-Test is performed for time shifts $q$ up to the order of ten. Up to this lag, the null hypothesis that the innovation process of returns is homoscedastic could not be rejected for most CCs (see Tab.\ \ref{TabelleStattest}),  i.e.\ the basic model $\epsilon_t = \sigma z_t$ with constant volatility $\sigma$ and $z_t$ being an independent and identical distributed process with mean 0 and variance 1 could not be rejected for the majority of CCs.
Note, that the results found in literature show a heterogeneous picture and we do not find ARCG effects in contrast to other related studies \citep{Peng.2018b, Dyhrberg.2016c, Avital.2014}. After all, only 13 of 28 time series examined show ARCH effects and only in 8 cases clearly. The differences in studies may derive from different sampling frequencies or the differently chosen time period or its length and show that the design of the data collection may have an influence on the results.
\begin{table}[H]
	\footnotesize \noindent
	\begin{center}
	\begin{tabularx}{0.75\textwidth}{l  r  r  r  r  r  r } \toprule
	CC								&	
	\multicolumn{3}{l}{ADF-Test}		& 
	\multicolumn{2}{l}{ARCH-Test}		&	
	Distribution						\\
	ID 									&	
	H0									&	
	Model								&	
	$p$-Value							&	
	H0							   		&	
	$p$-Value							&		
	{}									\\ 	\midrule
	EWCI$^-$	&	0	&	L	&	29.4	&	1	&	0.0	&	-	\\
	ANC	&	0	&	G	&	11.7	&	0	&	6.6	&	IID	\\
	BTB	&	0	&	L	&	22.9	&	1	&	0.0	&	-	\\
	BTC	&	0	&	G	&	48.7	&	1	&	3.1	&	$\approx$ IID	\\
	CSC	&	0	&	G	&	49.2	&	0	&	50.8	&	IID	\\
	DEM	&	0	&	G	&	24.0	&	1	&	0.0	&	-	\\
	DMD	&	0	&	G	&	75.8	&	0	&	18.0	&	IID	\\
	DGC	&	0	&	G	&	11.0	&	1	&	0.0	&	-	\\
	DOGE	&	0	&	L	&	14.7	&	0	&	69.8	&	IID	\\
	FTC	&	0	&	L	&	14.9	&	1	&	0.0	&	-	\\
	FLO	&	1	&	L	&	4.0	&	0	&	63.2	&	$\approx$ IID	\\
	FRC	&	0	&	L	&	14.3	&	0	&	71.6	&	IID	\\
	GLC	&	0	&	G	&	65.2	&	0	&	16.6	&	IID	\\
	IFC	&	0	&	G	&	6.4	&	0	&	38.3	&	IID	\\
	LTC	&	0	&	G	&	34.2	&	0	&	25.2	&	IID	\\
	MEC	&	0	&	G	&	17.7	&	1	&	1.5	&	$\approx$ IID	\\
	NMC	&	0	&	G	&	42.0	&	0	&	38.1	&	IID	\\
	NVC	&	0	&	G	&	22.8	&	0	&	92.6	&	IID	\\
	NXT	&	0	&	L	&	50.4	&	1	&	0.0	&	-	\\
	OMNI	&	0	&	G	&	35.7	&	0	&	70.2	&	IID	\\
	PPC	&	0	&	G	&	41.5	&	1	&	0.4	&	$\approx$ IID	\\
	XPM	&	0	&	G	&	12.7	&	0	&	11.1	&	IID	\\
	QRK	&	1	&	L	&	2.0	&	1	&	0.3	&	$\approx$ IID	\\
	XRP	&	0	&	L	&	35.1	&	1	&	0.0	&	-	\\
	TAG	&	0	&	G	&	6.0	&	0	&	60.4	&	IID	\\
	TRC	&	0	&	G	&	46.0	&	1	&	1.9	&	$\approx$ IID	\\
	WDC	&	1	&	L	&	2.5	&	1	&	0.0	&	-	\\
	ZET	&	0	&	G	&	22.6	&	0	&	6.0	&	IID	\\
	 \bottomrule
\end{tabularx}
\end{center}
	\caption{: Results of the statistical tests. Units in percent and boolean, see text.}
	\label{TabelleStattest}
\end{table}
The aforementioned results would justify the following calculation: $\text{E}[r_t] = \mu =c $ and $\text{Var}[r_t] = \sigma^2 $ for the corresponding CCs. Estimates for the mean and the variance of the individual returns are noted in Tab.\ \ref{TabelleBasicStatistik}. Their calculation is also justified with the combined consideration of the test results described above.

Combining both tests, we have noted a characterization of the return distribution in the last column of Tab.\ \ref{TabelleStattest}.
For the majority of CCs, independent and identical distributed (IID) returns can be assumed. 
Another part is approximately independent and identical distributed ($\approx$ IID), because either the lag structure is less important in the ADF model or the rejection of homoscedasticity based on the $p$-value is only weakly justified.
For eight CCs the assumption of IID returns is clearly rejected.

Note, that the test for independently identical ditstributed returns could have been carried out with a turning point test \citep{Bienayme.1874, Kendall.1977}. However, as we are interested in a deeper analysis of the possible serial correlation in our data sets, we use a combination of the augmented Dickey-Fuller test and the ARCH-Test instead.

All tests carried out so far, do not reject the assumption that the returns
obey an essentially symmetrical, unimodal distribution.
Furthermore, for the majority of CCs the assumption of IID or nearly IID holds.
In the first case (IID), the modeling of the empirical distribution with a distribution function is justified. In the second case ($\approx$ IID), the model represents a coarser approximation. In the latter case, if the assumption of IID was to be rejected, the distribution function can only be used as a rough approximation and must be
-- just like the case ($\approx$ IID) --  examined more precisely and critically in individual cases as we show in the following section. 

\subsection{Determination of the appropriate return distribution function} \label{sdiCC}
When modeling the empirical distribution of returns, we focus on families of unimodal distribution functions that are defined over the entire axis (infinite support). Hence, a more detailed investigation of the following distribution functions suggests itself: normal distribution (N), the generalized extremvalue distribution (GED), the generalized logistic distribution type 0 and type 3 (GLD0, GLD3) and the SDI. \begin{comment}which is also referred to as the L\'evy alpha-stable distributions \citep{Levy.1925}.\end{comment}
\begin{comment}
Note that the family of generalized extremvalue distributions, depending on the shape parameter, subsumes the special cases of the Gumbel, Weibull and Frechet distributions and is only defined for the Gumbel distribution over the entire axis, see e.g.\ \citet{Embrechts.1997}.

In a Finance context, the family of generalized logistic distributions has been analyzed by \citet{Fischer.2000}. Originating in early works of \citet{Gumbel.1944} and since then this family has a broad applications in various scientific fields, see e.g.\ \citet{Davidson.1980} and the literature cited therein. A more theoretical analysis of this family can be found in \citet{Nassar.2012}.
\end{comment}
The analysis below proves that the family of SDIs is the most promising for modeling the distribution of CC returns. 
Therefore, this family of functions will be introduced here in more detail. Several different parametrizations exist for the SDI. In the following formulation we follow the presentation and the SDI's parametrization described in \citet[Def.\ 1.4 therein]{Nolan.2020}.

SDIs represent a class of distributions suitable for modeling heavy tails and skewness. A linear combination of two independent, identically and stable distributed random variables has the same distribution as the individual variables.
A random variable $X$ follows the SDI $S(\alpha, \beta, \gamma, \delta)$
if its characteristic function is given by:
\begin{flalign}\label{SDI}\nonumber
& \text{E}\left[ \exp\left(\text{i}tX\right) \right]   = \\[8pt]
&
\begin{cases}
\exp\left(\text{i} \delta t - \left|\gamma t\right|^{\alpha}\Big[
1 + \text{i} \beta\text{sign}(t)\;\tan\left(\frac{\pi\alpha}{2}\right) \left(\left|  \gamma t\right|^{1-\alpha} - 1 \right)
\Big] \right)
& \alpha \neq 1 \\[8pt]
\exp\left(\text{i} \delta t - \left|\gamma t\right|\hphantom{{^{\alpha}}}\Big[
1 + \text{i} \beta\text{sign}(t)\;\frac{2}{\pi}
\ln\left(\left|  \gamma t\right| \right)
\Big]  \right) 
& \alpha = 1 
\end{cases}
\end{flalign}

The first parameter $0<\alpha\leq 2$ is called shape parameter and describes the tail of the distribution. This parameter may also be denoted as an {\it index of stability} or as a {\it characteristic exponent}.
The second parameter $-1\leq \beta \leq +1$ is a skewness parameter. If $\beta = 0$, then the distribution is symmetric otherwise it is left-skewed ($\beta<0$) or right-skewed ($\beta>0$). When $\alpha$ is small, the skewness of $\beta$ is significant. As $\alpha$ increases, the effect of $\beta$ decreases. Further, $\gamma\in\mathbb{R^+}$ is called the scale parameter and $\delta\in\mathbb{R}$ is the location parameter.

For the special case $\alpha = 2$ the characteristic function Eq.\ (\ref{SDI}) reduces to 
$\text{E}\left[ \exp\left(\text{i}tX\right) \right] = \exp\left(\text{i} \delta t - (\gamma t)^{2}\right)$ 
and becomes independent of the skewness parameter $\beta$ and
the SDI becomes equal to N with mean $\delta$ and standard deviation $\sigma = \sqrt{2}\gamma$.
\begin{comment}This is an important property for portfolio theory, for example, when considering multivariate distributions. Because it is basically possible to model normally distributed components of a random vector with the same function class.\end{comment}

\subsubsection*{Evaluation of distance measurements to compare model quality}
In the following we use standard distance measures to determine and compare the model qualities of N, GED, GLD0, GLD3 and SDI for the individual CCs.
There are several distance measures available which are suitable to measure the "discrepancy" between the empirical distribution function and a modelled distribution function. The distance measures from \citet{Cramer.1928} \citet{Mises.1931} ($W^2$-Distance), \citet{Anderson.1952, Anderson.1954} ($A^2$-Distance) and \citet{Kolmogorov.1933} \citet{Smirnov.1936, Smirnov.1948} (KS-Distance) are widely used in literature. A brief summary of the used distance measures are given in \ref{Appendix_TablesOfResults}.

In Tab.\ \ref{Tabelle_ADTest} we summarize the results for the Anderson Darling distance ($A^2$). The results for the Kolmogorov Smirnov distance (KS) and the Cram\'er von Mises distance ($W^2$) are compiled in Tab.\ \ref{Tabelle_KSTest} and Tab.\ \ref{Tabelle_CMTest} in \ref{Appendix_TablesOfResults}.

\begin{table}[h]
	\footnotesize \noindent
	\centering
	\begin{tabularx}{0.75\textwidth}{l  r  r  r  r  r  r} \toprule
		CC								&	
		\multicolumn{5}{l}{Anderson Darling Distance $A^2$} &
		Best Choice							\\
		ID 									&	
		N									&	
		GED									&	
		GLD0  								&	
		GLD3							   	&	
		SDI									&
		{}									\\ 	\midrule
		EWCI$^-$ &	3.41	&	147.4	&	1.76	&	0.81	&	0.79	&	SDI	\\
		ANC	&	8.65	&	22.4	&	2.53	&	1.61	&	0.39	&	SDI	\\
		BTB	&	3.36	&	12.3	&	0.76	&	0.48	&	0.22	&	SDI	\\
		BTC	&	2.07	&	224.8	&	0.88	&	0.60	&	0.97	&	GLD3	\\
		CSC	&	21.80	&	42.6	&	3.90	&	n.d.	&	0.43	&	SDI	\\
		DEM	&	2.64	&	6.7	&	0.54	&	0.17	&	0.20	&	GLD3	\\
		DMD	&	3.80	&	23.7	&	1.25	&	0.29	&	0.54	&	GLD3	\\
		DGC	&	6.84	&	26.6	&	2.05	&	0.65	&	0.78	&	GLD3	\\
		DOGE&	9.56	&	9.1	&	4.10	&	1.77	&	0.53	&	SDI	\\
		FTC	&	9.41	&	14.2	&	1.77	&	1.05	&	0.17	&	SDI	\\
		FLO	&	1.74	&	1.5	&	0.54	&	0.54	&	0.32	&	SDI	\\
		FRC	&	17.36	&	n.d.	&	5.21	&	1.98	&	0.39	&	SDI	\\
		GLC	&	1.53	&	16.8	&	0.40	&	0.19	&	0.42	&	GLD3	\\
		IFC	&	n.d.	&	n.d.	&	4.79	&	3.43	&	2.57	&	SDI	\\
		LTC	&	7.10	&	13.7	&	2.72	&	0.72	&	0.71	&	SDI	\\
		MEC	&	9.19	&	18.8	&	3.71	&	1.25	&	0.45	&	SDI	\\
		NMC	&	6.02	&	60.5	&	1.73	&	0.48	&	0.47	&	SDI	\\
		NVC	&	17.24	&	51.3	&	4.08	&	1.47	&	0.47	&	SDI	\\
		NXT	&	6.69	&	17.6	&	2.39	&	0.96	&	0.41	&	SDI	\\
		OMNI&	1.25	&	4.4	&	0.25	&	0.24	&	0.24	&	SDI	\\
		PPC	&	4.93	&	31.9	&	1.63	&	0.61	&	0.48	&	SDI	\\
		XPM	&	5.66	&	5.0	&	1.53	&	0.67	&	0.27	&	SDI	\\
		QRK	&	6.32	&	9.5	&	2.12	&	0.53	&	0.52	&	SDI	\\
		XRP	&	13.90	&	13.0	&	5.71	&	2.79	&	0.62	&	SDI	\\
		TAG	&	4.85	&	5.4	&	1.23	&	0.30	&	0.64	&	GLD3	\\
		TRC	&	4.36	&	5.5	&	0.80	&	0.45	&	0.13	&	SDI	\\
		WDC	&	9.68	&	24.1	&	3.95	&	1.35	&	0.57	&	SDI	\\
		ZET	&	5.23	&	12.3	&	1.58	&	0.41	&	0.48	&	GLD3	\\
		\bottomrule
	\end{tabularx}
	\caption{: Anderson Darling Distance for different body model distributions.}
	\label{Tabelle_ADTest}
\end{table}

The values of the various distance measures show that the GED is least suitable to model the empirical distribution function. This may derive from the fact that the GED contains a fundamental skewness, which can only be influenced slightly via a parameter selection. 
Further, as soon as the shape parameter becomes different from zero a fundamental change of the distribution model occurs and the definition interval on the $x$-axis becomes restricted.  
Additionally, associated with a change of sign of the shape parameter there is a fundamental change of the distribution model as well and an abrupt change of the sign of the upper (or lower) bound on the $x$-axis, see e.g.\ \citet{Embrechts.1997}. In the present case, these properties of the GED make it difficult to precisely adapt the distribution to the data set. 

On closer inspection of the calculated distances, the N does not appear to be suitable as a model either, since it is neither suitable for the modeling of empirical distribution functions with fat tails nor for those with a slight skew. Overall, the GLD0 shows significantly smaller distances across all CCs, but is also not ideally suited, as it is completely symmetrical and is therefore not able to model any slight skewness.
The best results in terms of the smallest distance can be achieved with the GLD3 and with the SDI. When comparing all CCs, the corresponding distances are very close to each other. If only the Cram\'er von Mises distance and the Kolmogorov Smirnov distance are considered, see \ref{Appendix_TablesOfResults}, it turns out that around half of the empirical distributions of CC returns can be modeled with one or the other distribution.
But as soon as more attention is paid to the tail, i.e.\ 
if deviations in the tail area are to be weighted more heavily in order to account for tail risks,
and the Anderson Darling distance $A^2$ is considered, the share of CCs, for which the SDI is the most suitable model, predominates.

This result ties in with the results of \citet{Majoros.2018, Kakinaka.2020}. Using intraday and daily time series for a small sample of CCs they show the SDI family to be the best choice for modeling the empirical distribution function of intraday and daily returns of CCs.

For a broader sample of CCs we found that the SDI is, on average, much better suited to model both slight skewness and pronounced tails in the empirical distribution function. Therefore we use the SDI for all CCs to model the distribution function of returns.\\

\begin{comment}This results tie in with results found in ... {\color{red} \cite{Gkillas.2018d}, \cite{Gkillas.2018} Gibt es hier weitere Literaturzitate A, B, C, ... die zu ähnlichen Ergebnissen kommen? Oder, Aussagen zu den anderen Verteilungen N, GED, GLD0+3, SDI machen?}\end{comment}

\subsubsection*{The stable distribution family as a model for cryptocurrencies}
Tab.\ \ref{TabelleBodyModel} shows the results of the parameter estimation for the SDI $S(\alpha, \beta, \delta, \gamma)$. 
\begin{table}[H]
	\footnotesize \noindent
	\begin{tabularx}{\textwidth}{lrrrr rrrrr r} \toprule
		CC								&	
		\multicolumn{8}{l}{Parameter of the SDI $S(\alpha, \beta, \delta, \gamma)$}	& 
		\multicolumn{2}{l}{AD-Test}		\\
		%%%%%
		ID 									&	
		$\hat\alpha$							&	
		$\pm\Delta\alpha$					&	
		$\hat\beta$								&
		$\pm\Delta\beta$ 					&
		$\hat\gamma$ 							&
		$\pm\Delta\gamma$ 					&
		$\hat\delta$ 							&
		$\pm\Delta\delta$ 					&
		H0							   		&	
		$p$-Value							\\ 	\midrule
		%%%%%
		EWCI$^-$	&	1.62	&	0.18	&	0.46	&	0.39	&	0.07	&	0.01	&	-0.01	&	0.02	&	0	&	48.9	\\
		ANC	&	1.41	&	0.17	&	0.26	&	0.31	&	0.16	&	0.02	&	-0.03	&	0.03	&	0	&	85.8	\\
		BTB	&	1.67	&	0.18	&	0.38	&	0.45	&	0.18	&	0.02	&	-0.03	&	0.04	&	0	&	98.3	\\
		BTC	&	1.78	&	0.16	&	-0.21	&	0.61	&	0.06	&	0.01	&	0.01	&	0.01	&	0	&	37.5	\\
		CSC	&	1.45	&	0.18	&	0.27	&	0.32	&	0.16	&	0.02	&	-0.04	&	0.03	&	0	&	81.4	\\
		DEM	&	1.65	&	0.18	&	-0.04	&	0.45	&	0.17	&	0.02	&	-0.01	&	0.03	&	0	&	99.0	\\
		DMD	&	1.56	&	0.18	&	0.10	&	0.39	&	0.11	&	0.01	&	-0.01	&	0.02	&	0	&	70.7	\\
		DGC	&	1.57	&	0.18	&	0.31	&	0.38	&	0.14	&	0.02	&	-0.04	&	0.03	&	0	&	49.7	\\
		DOGE&	1.33	&	0.17	&	0.35	&	0.27	&	0.08	&	0.01	&	-0.02	&	0.02	&	0	&	72.0	\\
		FTC	&	1.63	&	0.18	&	0.54	&	0.38	&	0.12	&	0.01	&	-0.05	&	0.03	&	0	&	99.7	\\
		FLO	&	1.88	&	n.d.	&	1.00	&	n.d.	&	0.16	&	n.d.	&	-0.02	&	n.d.	&	0	&	92.3	\\
		FRC	&	1.24	&	0.16	&	0.06	&	0.26	&	0.13	&	0.02	&	-0.02	&	0.03	&	0	&	86.0	\\
		GLC	&	1.80	&	0.16	&	0.43	&	0.63	&	0.15	&	0.02	&	-0.02	&	0.03	&	0	&	82.5	\\
		IFC	&	1.47	&	0.18	&	0.25	&	0.33	&	0.12	&	0.01	&	-0.04	&	0.02	&	1	&	4.6	\\
		LTC	&	1.37	&	0.17	&	0.06	&	0.30	&	0.06	&	0.01	&	-0.01	&	0.01	&	0	&	55.2	\\
		MEC	&	1.25	&	0.16	&	-0.01	&	0.26	&	0.09	&	0.01	&	-0.02	&	0.02	&	0	&	79.9	\\
		NMC	&	1.48	&	0.18	&	0.07	&	0.34	&	0.08	&	0.01	&	-0.01	&	0.02	&	0	&	78.0	\\
		NVC	&	1.42	&	0.17	&	0.28	&	0.31	&	0.08	&	0.01	&	-0.03	&	0.02	&	0	&	77.5	\\
		NXT	&	1.48	&	0.18	&	0.37	&	0.32	&	0.10	&	0.01	&	-0.03	&	0.02	&	0	&	83.8	\\
		OMNI&	1.82	&	0.16	&	0.26	&	0.71	&	0.14	&	0.01	&	-0.03	&	0.03	&	0	&	97.7	\\
		PPC	&	1.50	&	0.18	&	0.10	&	0.35	&	0.08	&	0.01	&	-0.02	&	0.02	&	0	&	76.8	\\
		XPM	&	1.58	&	0.18	&	0.42	&	0.37	&	0.10	&	0.01	&	-0.04	&	0.02	&	0	&	95.6	\\
		QRK	&	1.45	&	0.18	&	0.17	&	0.33	&	0.12	&	0.02	&	-0.03	&	0.02	&	0	&	72.6	\\
		XRP	&	1.33	&	0.17	&	0.35	&	0.27	&	0.07	&	0.01	&	-0.03	&	0.01	&	0	&	63.2	\\
		TAG	&	1.63	&	0.18	&	0.10	&	0.44	&	0.12	&	0.01	&	-0.02	&	0.02	&	0	&	61.1	\\
		TRC	&	1.64	&	0.18	&	0.29	&	0.43	&	0.13	&	0.02	&	-0.04	&	0.03	&	0	&	99.9	\\
		WDC	&	1.26	&	0.16	&	0.08	&	0.26	&	0.10	&	0.01	&	-0.03	&	0.02	&	0	&	67.2	\\
		ZET	&	1.53	&	0.18	&	0.15	&	0.36	&	0.13	&	0.02	&	-0.02	&	0.03	&	0	&	76.5	\\
		
		\bottomrule
	\end{tabularx}
	\caption{: Parameter of the SDI and goodness of fit test.}
	\label{TabelleBodyModel}
\end{table}
For the individual parameters of the SDI, the 95\% scatter intervals are provided, as well. The latter can be determined from the covariance matrix of the estimated parameters.
The covariance matrix of the parameter estimates is a matrix in which the off-diagonal element $(i, j)$ resembles the covariance between the estimates of the $i$-th parameter and the $j$-th parameter. For the CC FLO, these scatter intervals cannot be determined numerically using the data set at hand. This is due to the fact that the corresponding empirical distribution on the far right shows a very long, pronounced tail and is strongly skewed to the right. The estimated parameter $\hat\beta$ of the SDI takes the value of 1 accordingly, see Tab.\ \ref{TabelleBodyModel}.
It may be the case that the single right tail data point, i.e.\ the return in period 62, represents an outlier, which is difficult to determine and correct afterwards without any further knowledge. 

\begin{comment}
Therefore, the data set is as it is and should not be changed by winsorizing or censoring techniques, so we are currently including the data set of the currency FLO unchanged in the evaluation. 
\end{comment}

On average, the estimated parameter $\hat\alpha$ exceeds 1.5. With parameter $\alpha$ increasing to its limit value of 2.0, it can be seen that the distribution function becomes similar to the N and the skewness parameter ($\beta\neq 0$) becomes increasingly insignificant.
The parameters $\beta$ and $\alpha$ of the SDI are mutually dependent and, as described above, the meaning of $\beta$ decreases when $\alpha$ increases. So it is generally difficult to infer the skewness from the value $\beta$ alone. A relative comparison of the distributions with regard to the skewness is only possible if $\alpha$ has the same value.
Hence, for a more precise analysis of a distribution's skewness, other methods are necessary. In this manner, we used the SIG-Test as an example and note the results in Tab.\ \ref{TabelleBasicStatistik}.

\begin{figure}[htbp]
	%\centering
	\captionsetup{labelfont = bf, labelsep = none}
	\includegraphics[width=0.995\textwidth]{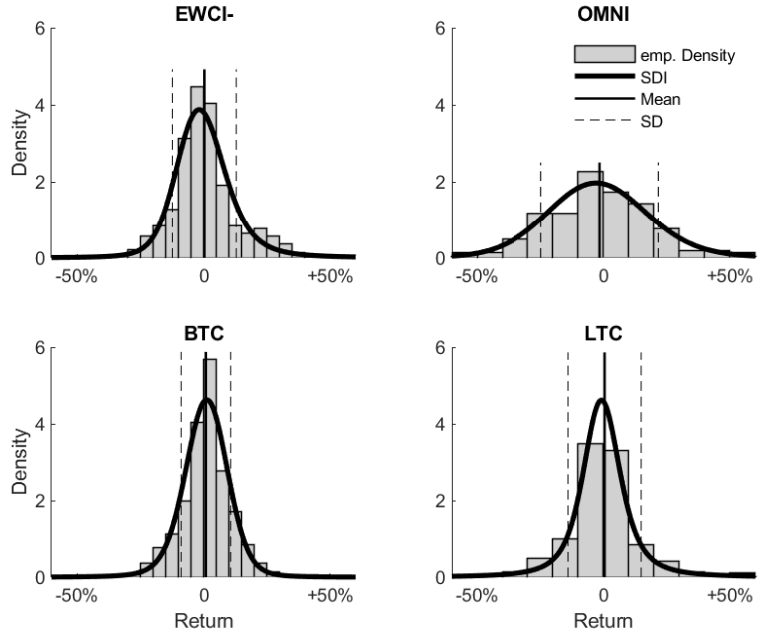}
	%\begin{quote}
	\caption[Histogram]{\label{Fig1Histogramme} The empirical densities as well as the results of the modeling of the distributions with the family of SDIs can be seen for the EWCI$^-$ and selected CCs.}
	%\end{quote}
\end{figure}

For some CCs and the EWCI$^-$ index, Fig.\ \ref{Fig1Histogramme} shows the empirical densities and the density of the corresponding SDI in comparison.

In addition, the complete goodness of fit test according to Anderson Darling was carried out. The last two columns of Tab.\ \ref{TabelleBodyModel} show that for almost all CCs with very high significance (high $p$-values), the null hypothesis that the adjusted SDI models the data set cannot be rejected. Only for the CC Infinitecoin (IFC) this assumption is rejected at the 5\% level. This is probably due to the fact that the empirical distribution suggests a slight bimodal distribution. This peculiarity of the CC Infinitecoin is also indicated in the results of the HDS-Test in Tab.\ \ref{TabelleBasicStatistik}. 

Overall, it can be stated that the SDI family represents a suitable framework for modeling the distribution function of CC returns. We will make use of this finding for the assessment of tail risks and the comparison of our results with other modeling approaches.

\section{Assessment of tail risks}\label{tailrisks}

\subsection{Modeling of cryptocurrencies' tail risks}\label{tailmodel}
Especially when considering high quantiles in the risk assessment process, separate modeling of the tail of the often unknown parent distribution is necessary. In practice, the GPD is used predominantly as a tail model \citep{BaselCommiteeonBankingSupervision.2009}. 
Henceforth, we briefly discuss the main steps of the tail modelling using the GPD in this section.

A theorem in extreme value theory, which goes back to \citet{Gnedenko.1943}, \citet{Balkema.1974} and \citet{PickandsIII.1975}, states that for a broad class of distributions, the distribution of the excesses over a threshold $u$ converges to a GPD if the threshold is sufficiently large.

The GPD is usually expressed as a two-parameter distribution and follows the distribution function:
\begin{flalign}\label{GPDDF}
F(x) & = 1-\left(1+\xi\frac{x}{\sigma} \right)^{-\frac{1}{\xi}},
\end{flalign}
where $\sigma$ is a positive scale parameter and $\xi$ is a shape parameter (sometimes called the {\it tail parameter}). The density function is described as
\begin{flalign}\label{GPDPF}
f(x) & = \frac{1}{\sigma}\left(1+\xi\frac{x}{\sigma} \right)^{-\frac{1+\xi}{\xi}},
\end{flalign}
with support $0\leq x< \infty$ for $\xi\geq 0$ and $0\leq x\leq -\frac{\sigma}{\xi}$ when $\xi< 0$. The mean and variance are depicted as ${\text E}[x] = \frac{\sigma}{1-\xi}$ and ${\text{Var}}[x] = \frac{\sigma^2}{(1-\xi)^2(1-2\xi)} $, respectively. Thus, the mean and variance of the GPD are positive and finite only for $\xi < 1$ and $\xi < 0.5$, respectively. For special values of $\xi$, the GPD leads to various other distributions. When $\xi = 0, -1$, the GPD translates to an exponential or a uniform distribution. For $\xi>0$, the GPD has a long tail to the right and resembles a reparameterized version of the usual Pareto distribution. Several areas of applied statistics have used the latter range of $\xi$ to model data sets that exhibit this form of a long tail. 

Since the GPD was introduced by \citet{PickandsIII.1975}, numerous theoretical advancements and applications have followed (\cite{Davison.1984}, \cite{Smith.1984}, \cite{Smith.1985}, \cite{vanMontfort.1985}, \cite{Hosking.1987}, \cite{Davison.1990}, \cite{Choulakian.2001}). Its applications include the use in analysis of extreme events in hydrology, as a failure-time distribution in reliability studies and in the modeling of large insurance claims. Numerous examples of applications can be found in \citet{Embrechts.1997} and the studies listed therein. The GPD is also increasingly used in financial and banking sectors. Especially in the assessment of risks based on high quantiles, the GPD is one of the proposed distributions for modeling the tail of an unknown parent distribution \citep{BaselCommiteeonBankingSupervision.2009}.

The preferred method in literature for estimating the parameters of the GPD is the standard maximum likelihood method (\cite{Davison.1984}, \cite{Smith.1984}, \cite{Smith.1985}, \cite{Hosking.1987}). \citet{Choulakian.2001} state that it is theoretically possible to have data sets for which no solution to the likelihood equations exists, however this appears to be an extremly rare case in practice.
In fact, in our examination of the 27 CCs and the EWCI$^-$, maximum likelihood estimates of the parameters could be deduced in every case.
Nevertheless, a plausibility check of the results is highly recommended. As we show in Sec.\ \ref{tailrisksassessment}, when assessing tail risks, it is advisable to evaluate the results, in order to avoid possible misinterpretations in single individual cases.

For the separate modeling of the tail, solely the parameters of the GPD need to be determined from the data pertaining to the tail region of the underlying empirical distribution function, i.e.\ the data which belong to the area below a threshold $u$ in case of the loss tail. \begin{comment} Or equivalently above threshold $u$, if the losses are represented positively by a change in sign.\end{comment} The correct determination of the threshold value $u$ is therefore essential. Various methods for determining the optimal threshold value are conceivable. In this paper, we use a recently developed fully automatic process that does not require any user intervention or additional parameters, following\ \citet{Hoffmann.2020, Hoffmann.2021}. A brief description of the procedure is given in \ref{Appendix_TailModel}.

Tab.\ \ref{TabelleTailModel} depicts the estimated parameters for the GPD for the different CCs. The second column gives the proportion of the whole dataset belonging to the loss tail. The proportion of the return data below the threshold $\hat u$ is used to fit the parameters $\xi, \sigma$ of the GPD. The threshold value lies within the bandwidth shown in Tab.\ \ref{TabelleBasicStatistik} and when considering the loss tail closer to the lower interval limit of the bandwidth. 
Due to the standard goodness of fit tests according to Cr\'amer von Mises (CvM) or Anderson Darling (AD), the null hypothesis that the GPD is a suitable model for the tail, can not be rejected on any significance level.
In addition, the $p$-values for the lower tail statistics (LT) according to \citet{Ahmad.1988} are given in Tab.\ \ref{TabelleTailModel}. The corresponding statistics $AL^2$ is defined in \ref{Appendix_TailModel} and used here to determine the threshold value $u$. As can be seen in column six of Tab.\ \ref{TabelleTailModel} the lower tail statistics show high confidence levels, as well.

\begin{table}[H]
	\footnotesize \noindent
	\centering
	\begin{tabularx}{0.720\textwidth}{lrrrr rrr} \toprule
		CC										&
		Prop.\										&	
		\multicolumn{3}{l}{GPD-Parameter}	&
		\multicolumn{3}{l}{Goodness of Fit} 	\\[5pt]
		%%%%%
		ID 											&
		{} 											&
		\multicolumn{1}{l}{$\hat u$} 				&
		\multicolumn{1}{c}{$\hat \xi$} 				&
		\multicolumn{1}{r}{$\hat \sigma$}			&							
		\multicolumn{3}{c}{$p$-Values}		\\ [5pt]
		\multicolumn{5}{l}{{ }} 			&
		\multicolumn{1}{l}{LT} 				&
		\multicolumn{1}{c}{CvM} 			&
		\multicolumn{1}{r}{AD}				\\	\midrule
		EWCI$^-$	&	45.7	&	-1.7	&	-0.09	&	0.09	&	88.0	&	84.9	&	92.6	\\
		ANC	&	4.3	&	-52.0	&	-0.08	&	0.61	&	98.5	&	98.1	&	93.9	\\
		BTB	&	4.3	&	-51.0	&	0.63	&	0.13	&	99.0	&	96.1	&	98.1	\\
		BTC	&	24.1	&	-3.9	&	-0.30	&	0.10	&	93.0	&	97.5	&	98.1	\\
		CSC	&	9.9	&	-35.5	&	0.82	&	0.11	&	96.7	&	99.7	&	94.3	\\
		DEM	&	51.4	&	-1.3	&	-0.05	&	0.22	&	54.2	&	58.2	&	60.8	\\
		DMD	&	4.3	&	-32.5	&	0.04	&	0.13	&	94.5	&	89.0	&	93.2	\\
		DGC	&	10.3	&	-31.7	&	0.61	&	0.08	&	99.8	&	99.6	&	99.5	\\
		DOGE&	21.6	&	-9.2	&	0.13	&	0.09	&	99.1	&	99.4	&	98.9	\\
		FTC	&	3.5	&	-37.0	&	0.99	&	0.05	&	99.7	&	99.4	&	93.7	\\
		FLO	&	16.3	&	-20.3	&	-0.25	&	0.17	&	93.9	&	91.5	&	89.2	\\
		FRC	&	11.0	&	-28.8	&	0.28	&	0.31	&	98.8	&	98.3	&	99.6	\\
		GLC	&	51.8	&	0.2	&	-0.19	&	0.20	&	99.9	&	99.9	&	100.0	\\
		IFC	&	20.9	&	-18.2	&	0.55	&	0.07	&	55.0	&	70.4	&	52.8	\\
		LTC	&	44.3	&	-0.7	&	-0.18	&	0.11	&	57.5	&	67.7	&	54.1	\\
		MEC	&	56.4	&	0.0	&	0.16	&	0.12	&	99.8	&	99.2	&	93.3	\\
		NMC	&	46.1	&	-1.9	&	0.15	&	0.09	&	78.0	&	95.7	&	94.0	\\
		NVC	&	9.2	&	-19.3	&	0.72	&	0.05	&	77.2	&	93.6	&	86.0	\\
		NXT	&	2.8	&	-34.6	&	1.27	&	0.03	&	60.1	&	76.7	&	67.7	\\
		OMNI&	15.2	&	-23.0	&	-0.11	&	0.14	&	95.9	&	96.2	&	97.1	\\
		PPC	&	8.5	&	21.3	&	0.00	&	0.09	&	96.7	&	97.1	&	98.8	\\
		XPM	&	4.6	&	29.5	&	-0.08	&	0.12	&	91.4	&	93.1	&	96.7	\\
		QRK	&	63.5	&	-1.6	&	0.05	&	0.16	&	47.7	&	60.8	&	61.1	\\
		XRP	&	2.8	&	-25.6	&	0.76	&	0.04	&	99.2	&	98.8	&	99.1	\\
		TAG	&	53.2	&	-0.5	&	-0.13	&	0.17	&	91.2	&	92.7	&	96.8	\\
		TRC	&	35.5	&	-9.9	&	-0.06	&	0.15	&	64.2	&	70.9	&	81.0	\\
		WDC	&	62.8	&	0.8	&	0.18	&	0.13	&	94.8	&	94.4	&	88.9	\\
		ZET	&	20.9	&	-17.8	&	0.27	&	0.11	&	59.8	&	83.6	&	82.5	\\
		\bottomrule
	\end{tabularx}
	\caption{: Parameter of the GPD and goodness of fit test for the loss tail. Units in percent.}
	\label{TabelleTailModel}
\end{table}

\subsection{Risk assessment at high quantiles}\label{tailrisksassessment}
In this section, we use the SDI as the body model and the GPD as the tail model to determine the risk parameters Value at Risk (as a quantile) and the Conditional Value at Risk (as a weighted loss when the loss threshold is exceeded), see i.a.\ \citet{Embrechts.1997,Hull.2018}. The data set includes $T = 282$ return observations for each CC, so that a comparison of the results with the empirically determined values is possible for moderately large confidence levels ($\approx 99\%$). 
The corresponding values for the confidence level of 99.9\%, which is important for regulatory purposes \citep{BaselCommiteeonBankingSupervision.2004, EuropeanParliament.2009, EuropeanParliament.2013b, EuropeanParliament.2013a}, can only be estimated for data records of this length using a previously fitted body or tail model.
The calculation of the quantiles is also subject to a statistical spread and the estimation error increases the fatter the tail is and the higher the confidence level is selected, see e.g.\ \citet{Hoffmann.2020b}.

\begin{comment}
The analysis and the estimation of the corresponding error band of quantiles goes far beyond the scope of this work and is not considered further here, but the latter must be considered in a risk report for regulatory purposes for the complete classification of the tail risk.
\end{comment}

\subsubsection*{Value at Risk}\label{valueatrisk}
Tab.\ \ref{TabelleVaRCalculation} illustrates the results of the risk assessment for the most common confidence levels found in literature and regulatory requirements. The Value at Risk for the observed CCs are shown for the various model. 
\begin{table}[H]
	\footnotesize \noindent
	\centering
	\begin{tabularx}{\textwidth}{l|rrrr|rrrr|rrrr} \toprule
		CC										&
		\multicolumn{12}{l}{Value at Risk} 				\\
		%%%%%
		{}											&
		\multicolumn{4}{l|}{empirical} 				& 
		\multicolumn{4}{l|}{Tail-Modell (GPD)} 		& 		
		\multicolumn{4}{l}{Body-Modell (SDI)} 		\\
		%%%%%
		ID 											&
		\multicolumn{1}{l}{{\tiny 95\%}} 			&
		\multicolumn{1}{l}{{\tiny 97\%}} 			&
		\multicolumn{1}{l}{{\tiny 99\%}}		 	&		
		\multicolumn{1}{r|}{{\tiny 99.9\%}}		 	&		
		\multicolumn{1}{l}{{\tiny 95\%}} 			&
		\multicolumn{1}{l}{{\tiny 97\%}} 			&
		\multicolumn{1}{l}{{\tiny 99\%}}		 	&		
		\multicolumn{1}{r|}{{\tiny 99.9\%}}		 	&				
		\multicolumn{1}{l}{{\tiny 95\%}} 			&
		\multicolumn{1}{l}{{\tiny 97\%}} 			&
		\multicolumn{1}{l}{{\tiny 99\%}}		 	&		
		\multicolumn{1}{r}{{\tiny 99.9\%}}		 	\\ \midrule			
		%%%%%		
		EWCI-	&	26	&	29	&	38	&	41	&	19	&	23	&	30	&	43	&	18	&	22	&	33	&	116	\\
		ANC	&	101	&	133	&	212	&	241	&	42	&	73	&	136	&	252	&	46	&	60	&	119	&	582	\\
		BTB	&	75	&	91	&	154	&	201	&	49	&	56	&	82	&	251	&	47	&	55	&	82	&	276	\\
		BTC	&	21	&	24	&	28	&	30	&	16	&	19	&	24	&	31	&	15	&	18	&	28	&	91	\\
		CSC	&	117	&	166	&	424	&	698	&	46	&	58	&	110	&	595	&	47	&	61	&	116	&	535	\\
		DEM	&	72	&	84	&	100	&	101	&	51	&	61	&	82	&	122	&	48	&	60	&	100	&	367	\\
		DMD	&	44	&	52	&	72	&	85	&	30	&	37	&	52	&	85	&	30	&	38	&	69	&	280	\\
		DGC	&	65	&	82	&	153	&	217	&	39	&	46	&	71	&	229	&	40	&	49	&	81	&	314	\\
		DOGE	&	36	&	43	&	56	&	61	&	23	&	29	&	42	&	77	&	23	&	31	&	64	&	343	\\
		FTC	&	50	&	60	&	109	&	145	&	36	&	38	&	50	&	205	&	32	&	37	&	53	&	177	\\
		FLO	&	49	&	54	&	66	&	68	&	38	&	44	&	55	&	70	&	37	&	42	&	52	&	68	\\
		FRC	&	110	&	142	&	256	&	332	&	57	&	78	&	136	&	333	&	54	&	79	&	185	&	1159	\\
		GLC	&	49	&	55	&	68	&	74	&	38	&	44	&	56	&	74	&	36	&	42	&	56	&	140	\\
		IFC	&	61	&	77	&	114	&	126	&	34	&	43	&	75	&	251	&	35	&	45	&	82	&	365	\\
		LTC	&	27	&	31	&	34	&	34	&	20	&	24	&	31	&	41	&	21	&	30	&	62	&	321	\\
		MEC	&	59	&	71	&	99	&	113	&	37	&	46	&	70	&	137	&	38	&	56	&	128	&	787	\\
		NMC	&	42	&	50	&	92	&	110	&	27	&	34	&	51	&	97	&	25	&	33	&	63	&	284	\\
		NVC	&	47	&	63	&	141	&	236	&	23	&	28	&	46	&	188	&	24	&	31	&	59	&	278	\\
		NXT	&	42	&	48	&	78	&	84	&	33	&	34	&	42	&	210	&	27	&	33	&	59	&	258	\\
		OMNI	&	48	&	55	&	67	&	73	&	37	&	43	&	55	&	76	&	37	&	42	&	56	&	142	\\
		PPC	&	35	&	40	&	55	&	60	&	26	&	31	&	41	&	61	&	25	&	33	&	60	&	258	\\
		XPM	&	40	&	46	&	62	&	68	&	29	&	35	&	47	&	70	&	27	&	33	&	51	&	191	\\
		QRK	&	62	&	71	&	91	&	94	&	41	&	50	&	70	&	117	&	38	&	50	&	96	&	440	\\
		XRP	&	30	&	36	&	57	&	73	&	24	&	25	&	32	&	85	&	22	&	30	&	61	&	330	\\
		TAG	&	47	&	53	&	62	&	64	&	35	&	41	&	53	&	73	&	34	&	42	&	69	&	252	\\
		TRC	&	51	&	57	&	69	&	73	&	37	&	44	&	57	&	83	&	36	&	44	&	68	&	238	\\
		WDC	&	64	&	79	&	109	&	122	&	39	&	50	&	76	&	150	&	40	&	57	&	128	&	769	\\
		ZET	&	59	&	73	&	91	&	98	&	37	&	46	&	70	&	151	&	37	&	47	&	85	&	357 \\		
		\bottomrule
	\end{tabularx}
	\caption{: Value at Risk of the CCs for different confidence levels and different calculation methods. Losses with a positive sign and units in percent.}
	\label{TabelleVaRCalculation}
\end{table}
Overall, it can be seen that in the overwhelming number of individual cases, the assessment of risk with the tail model (GPD) is closer to the empirical Value at Risk values. This applies to the lower confidence levels in particular, but even with high confidence level of 99.9\%, better estimates are possible in individual cases compared to the body model. 
This becomes apparent from the statistical parameters of the deviation analysis shown in Tab.\ \ref{TabelleVaRCompare} as well. The mean value and standard deviation over the set of CCs are shown. For this purpose, the deviation between the modeled variable and the corresponding empirical Value at Risks was determined. It turns out that, on average, adjusting the GPD as tail model leads to better risk estimates. 

\begin{table}[H]
	\footnotesize \noindent
	\centering
	\begin{tabularx}{0.65\textwidth}{lrcl rrrr} \toprule
		{} 												&   
		\multicolumn{3}{l}{$\Delta$VaR	}				&
		\multicolumn{4}{l}{Confidence levels} 		    \\
		%%%%%
		{} 												&
		{}												&
		{}												&
		{}												&											
		\multicolumn{1}{l}{{\tiny 95\%}} 				&
		\multicolumn{1}{l}{{\tiny 97\%}} 				&
		\multicolumn{1}{l}{{\tiny 99\%}}		 		&		
		\multicolumn{1}{r}{{\tiny 99.9\%}}		 		\\ \midrule			
		%%%%%		
		Mean	&	GPD & ./. & Emp.\	&	0.5	    &	-0.2	&	-4.6	&	15.5	\\
				&	SDI & ./. & Emp.\	&	-0.3	&	0.4	    &	10.5	&	214.1	\\[5pt]
		SD		&	GPD & ./. & Emp.\	&	2.4	    &	2.1	    &	7.3	    &	43.1	\\
				&	SDI & ./. & Emp.\	&	1.8	    &	4.1	    &	19.3	&	211.7	\\
		\bottomrule
	\end{tabularx}
	\caption{: Average deviation from the empirical Value at Risk and scattering.}
	\label{TabelleVaRCompare}
\end{table}
When comparing CCs with each other, a heterogeneous picture emerges, see Tab.\ \ref{TabelleVaRCalculation}. If the empirical Value at Risk for the 99.9\% confidence interval is taken as a measure, two subgroups can be defined for the cut-off value $\text{VaR}_{99.9\%} \approx$ 100\%. One group possesses a significantly lower tail risk ($\text{VaR}_{99.9\%} <$ 100\%) as the corresponding $\hat\xi$ values in Tab.\ \ref{TabelleTailModel} indicate. The other group ($\text{VaR}_{99.9\%} >$ 100 \%) partly embodies a significantly higher tail risk. Correspondingly, large $\hat\xi$ values can be determined for these CCs. \\

\begin{figure}[htbp]
	%\centering
	\captionsetup{labelfont = bf, labelsep = none}
	\includegraphics[width=0.995\textwidth]{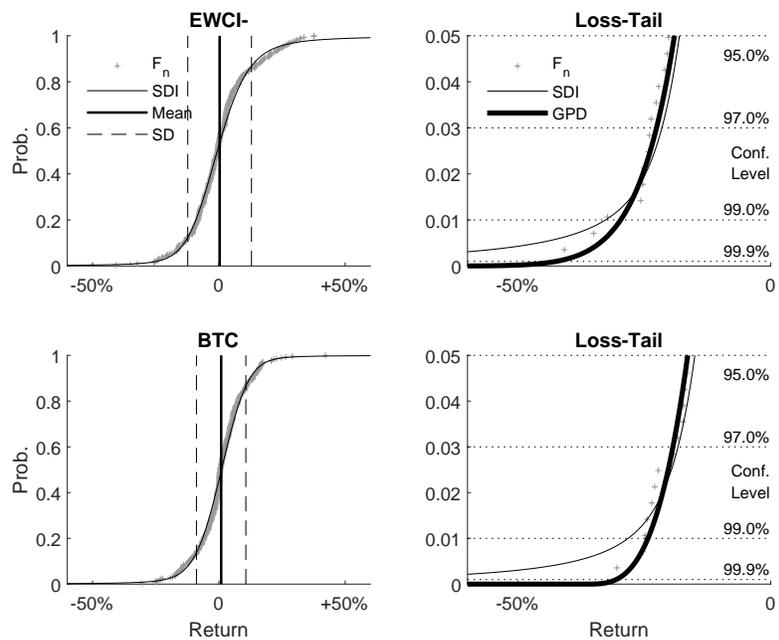}
	%\begin{quote}
	\caption[Probability function and high quantiles]{\label{Fig2ProbFktQuantile} The empirical distribution function and the distribution function modeled with the SDI can be seen for the EWCI$^-$ and an exemplary selected CC (left panels). The right graphics focus on the loss tail. The GPD adapted as a tail model and the confidence levels that are important for the regulator are also shown.}
	%\end{quote}
\end{figure}

Fig.\ \ref{Fig2ProbFktQuantile} shows the empirical distribution function for the EWCI$^-$ and Bitcoin (BTC), the SDI as a body model and the GPD as a tail model in comparison. The graphics on the right portray an enlargement of the loss area. Particularly in this region, the GPD models the empirical distribution function very well. 
Considering the afore conducted analysis, it can be stated that the GPD is ideally suited to carry out risk assessment at high quantiles. Therefore, we exclusively consider the GPD to estimate the Conditional Value at Risk as a further risk indicator in the following.

\subsubsection*{Conditional Value at Risk}\label{conditionalvalueatrisk}
The following Tab.\ \ref{TabelleCVaRCalculation} shows the Conditional Value at Risk calculated with the tail model (see Tab.\ \ref{TabelleTailModel}) for the individual CCs.
The calculation of the Conditional Value at Risk can be conducted using the mean excess function of the GPD:
\begin{flalign}\label{MeanExcessGPD}
e(v) & = \frac{\sigma + \xi v}{1 - \xi},
\end{flalign}
with $v$ greater than the lower bound of the definition interval of the GPD considering the loss tail and the parameters $\sigma, \xi$ of the GPD. The mean excess function is bound to the restrictions: $\xi<1$, and $\sigma + \xi v > 0$, see e.g.\ \citet[Theorem 3.4.13]{Embrechts.1997}.
In Tab. \ref{TabelleCVaRCalculation} we used Eq.\ (\ref{MeanExcessGPD}) to estimate the Conditional Value at Risk for each CCs setting $v = \text{VaR}_{p\%}$.

Once more, the grouping of CCs described above can be seen. A group of CCs with a fat tail and therefore higher tail risk can be distinguished from a group with moderate risk, see Tab.\ \ref{TabelleCVaRCalculation}.
\begin{table}[H]
	\footnotesize \noindent
	\centering
	\begin{tabularx}{0.460\textwidth}{lrrrr} \toprule
		CC										&
		\multicolumn{4}{l}{Conditional Value at Risk} \\
		%%%%%
		ID 											&
		\multicolumn{1}{l}{{\tiny 95\%}} 			&
		\multicolumn{1}{l}{{\tiny 97\%}} 			&
		\multicolumn{1}{l}{{\tiny 99\%}}		 	&		
		\multicolumn{1}{r}{{\tiny 99.9\%}}		 	\\ \midrule			
		%%%%%		
		EWCI-	&	25	&	29	&	35	&	47	\\
		ANC	&	96	&	125	&	184	&	292	\\
		BTB	&	169	&	188	&	258	&	717	\\
		BTC	&	20	&	23	&	26	&	31	\\
		CSC	&	308	&	374	&	655	&	3294	\\
		DEM	&	70	&	79	&	99	&	138	\\
		DMD	&	45	&	52	&	68	&	102	\\
		DGC	&	118	&	136	&	200	&	600	\\
		DOGE&	37	&	43	&	59	&	98	\\
		FTC	&	3219	&	3408	&	4343	&	16648	\\
		FLO	&	44	&	49	&	58	&	69	\\
		FRC	&	122	&	151	&	231	&	504	\\
		GLC	&	49	&	54	&	64	&	79	\\
		IFC	&	91	&	112	&	181	&	571	\\
		LTC	&	26	&	29	&	35	&	44	\\
		MEC	&	59	&	70	&	99	&	178	\\
		NMC	&	43	&	51	&	71	&	125	\\
		NVC	&	99	&	115	&	180	&	679	\\
		NXT	&	-134	&	-138	&	-165	&	-784	\\
		OMNI&	46	&	51	&	62	&	80	\\
		PPC	&	35	&	40	&	49	&	70	\\
		XPM	&	38	&	43	&	55	&	76	\\
		QRK	&	59	&	69	&	90	&	140	\\
		XRP	&	114	&	121	&	147	&	367	\\
		TAG	&	46	&	52	&	62	&	80	\\
		TRC	&	49	&	55	&	68	&	92	\\
		WDC	&	63	&	76	&	107	&	197	\\
		ZET	&	66	&	79	&	112	&	223	\\
		\bottomrule
	\end{tabularx}
	\caption{: Conditional Value at Risk of CCs for different confidence levels calculated with the corresponding tail model (GPD). Losses with a positive sign and units in percent.}
	\label{TabelleCVaRCalculation}
\end{table}
Furthermore, two peculiarities are noticeable concerning the CCs Feathercoin (FTC) and Nxt (NXT). For both CCs, the tail is modeled on a small number of data points which have been assigned to the tail. This individual property of the data set deriving from the random distribution of the data in the tail area, is assumed to be given and, as already noted, is not corrected. In particular, when estimating the parameter $\xi$, small samples lead to large statistical errors. Regrading the conspicuous CCs, it can be observed that the parameter is very close to 1 in one case (FTC) and even higher in the other case (NXT), see Tab.\ \ref{TabelleTailModel}. As a result, the calculation of the Conditional Value at Risk for the CC NXT is not possible and must be discarded, cf.\ Eq.\ (\ref{MeanExcessGPD}) and the restriction $\xi <1$. On the other hand, the calculation of the FTC with $\xi\approx 1$ has to be questioned critically. Hence, the Conditional Value at Risk may only be a rough estimate in this case. \begin{comment}These are observations that are of importance in practice, from which the regulatory obligation to check the plausibility of individual results follows.\end{comment}
%
\section{Conclusion}
\begin{comment}The latter can be used for portfolio optimization as a representative of the CC asset class as part of a strategic asset allocation.
The quantitative portfolio optimization can only be carried out if there is knowledge of the underlying stochastic process from theoretical considerations or if a suitable distribution model is available. A theoretical distribution derived from fundamental considerations about CCs seems difficult to obtain, which is why we investigate which distribution model is best suited to describe our empirical data.\end{comment}
This study aims to find a distribution which models CC returns most accurateley and does not suffer from restrictions in specific parts of the distribution. In former research the SDI and GPD have been found to model the body or the tail of the CC return distributions adequately respectively. Nevertheless both distributions prove to be unsuitable to model the entirety of the distrubition appropriately. Therefore, using a novel approach to separate of distribution's tail from its body, we model the entire distribution by combining the model abilities of the SDI for the body and the GPD for the tail.  

We select 27 CCs from the broad market of CCs according to predefined criteria and construct the representative index EWCI$^-$. Overall, we find independent, identical distribution like the GPD and the SDI to be well suited for the most part and the family of SDIs in particular to be able to model the slightly skewed empirical distributions especially in the body region. A comparison between different distribution functions shows that the SDI has outstanding modeling properties across the entire data set. However, we show that the assessment of risks associated with fat tails can be carried out more precisely with the GPD.
The analysis of tail risks in the CC market using the GPD further hints towards a certain internal structure of the CC market. It turns out that the CC market can roughly be divided into two sets: CCs with moderate risk and CCs with high risk. This finding provides valuable information for both investors and regulators alike.
Hence, our results are not only relevant for scientific applications and extensions but for conceivable future regulation as well, if the asset class of CC is to become a permanent, noteworthy component of institutional investors' portfolios in the financial sector in the future.
In this regard, numerous future extensions and topics of research are conveivable. On the one hand, the SDI's suitability to model CC returns in portfolio optimization remains to be investigated. On the other hand, further research considering the segmentation of the CC market could enrich the understanding of CCs and improve forecasts concerned with the fundamental behavior of different CCs.

\appendix
\section{Distance measures -- Tables of results}\label{Appendix_TablesOfResults}
\subsection{Distance measures}\label{Appendix_DistanceMeasures}
In what follows a brief summary of the used distance measures is given.
\subsubsection*{Cram\'er von Mises and Anderson Darling distance measures}
As a convenient measure of the distance or "discrepancy" between the empirical distribution functions $F_n(x)$ due to
\citet{Kolmogorov.1933} and a model $F(x)$, usually the weighted mean square error is considered
\begin{flalign}\label{WMSEEQ}
\hat R_n & = n \int_{-\infty}^{+\infty} \left(F_n(x)-F(x) \right)^2\; w(F(x))\;\diff F(x),
\end{flalign}
which was introduced in the context of statistical test procedures by \citet{Cramer.1928}, \citet{Mises.1931} and \citet{Smirnov.1936}; compare also \citet{Shorack.2009} and the references therein. In decision theory \citep{Ferguson.1967}, the weighted mean square error has broad applicability in the determination of the unknown parameters of distributions via minimum distance methods \citep{Wolfowitz.1957, Blyth.1970, Parr.1980, Boos.1982}.
This measure of error is also used when adapting tail models \citep{Hoffmann.2020, Hoffmann.2021}. Therefore, in Sec.\ \ref{tailmodel} we applied this error measure in connection with the adaption of a suitable tail model for CC returns. A brief overview of the procedure to fit a suitable tail model is given in \ref{Appendix_TailModel}.

The non-negative weight function $w(t)$ in Eq.\ (\ref{WMSEEQ}) is a suitable preassigned function for accentuating the difference between the distribution functions in the range where the distance measure is desired to have sensitivity. Consider the weight function
\begin{flalign}\label{WFEQ}
w(t) & = \frac{1}{t^a(1-t)^b}
\end{flalign}
for real-valued stress parameters $a,b\geq 0$ and $t\in[0,1]$. Here, $a$ affects the weight at the lower tail and $b$ at the upper tail. Then, for $a=b=0$, Eq.\ (\ref{WMSEEQ}) provides the Cram\'er von Mises distance $W^2$ used in the corresponding statistic \citep{Cramer.1928, Mises.1931}, whereas when heavily weighting the tails ($a=b=1$), it is equal to the Anderson Darling distance $A^2$ used in the corresponding statistic \citep{Anderson.1952, Anderson.1954}. The Anderson Darling distance weights the difference between the two distributions simultaneously more heavily at both ends of the distribution $F(x)$.

\subsubsection*{Cram\'er von Mises Distance}\label{Appendix_CMDistance}
\begin{table}[H]
	\footnotesize \noindent
	\centering
	\begin{tabularx}{0.735\textwidth}{l  r  r  r  r  r  r} \toprule
		CC								&	
		\multicolumn{5}{l}{Cram\'er von Mises Distance $W^2$} &
		Best Choice							\\
		ID 									&	
		N									&	
		GED									&	
		GLD0  								&	
		GLD3							   	&	
		SDI									&
		{}									\\ 	\midrule
		EWCI$^-$&	0.62	&	29.89	&	0.23	&	0.07	&	0.12	&	GLD3	\\
		ANC	&	1.51	&	4.25	&	0.41	&	0.24	&	0.06	&	SDI	\\
		BTB	&	0.55	&	2.22	&	0.11	&	0.06	&	0.04	&	SDI	\\
		BTC	&	0.40	&	41.81	&	0.16	&	0.11	&	0.19	&	GLD3	\\
		CSC	&	3.94	&	8.22	&	0.60	&	0.15	&	0.09	&	SDI	\\
		DEM	&	0.43	&	1.20	&	0.08	&	0.02	&	0.04	&	GLD3	\\
		DMD	&	0.70	&	5.19	&	0.22	&	0.05	&	0.10	&	GLD3	\\
		DGC	&	1.21	&	5.33	&	0.33	&	0.06	&	0.14	&	GLD3	\\
		DOGE&	1.86	&	1.70	&	0.64	&	0.20	&	0.07	&	SDI	\\
		FTC	&	1.54	&	2.37	&	0.21	&	0.09	&	0.03	&	SDI	\\
		FLO	&	0.29	&	0.23	&	0.07	&	0.07	&	0.05	&	SDI	\\
		FRC	&	3.26	&	6.35	&	0.94	&	0.33	&	0.07	&	SDI	\\
		GLC	&	0.27	&	3.66	&	0.06	&	0.02	&	0.08	&	GLD3	\\
		IFC	&	2.92	&	4.43	&	0.91	&	0.72	&	0.58	&	SDI	\\
		LTC	&	1.35	&	2.98	&	0.46	&	0.11	&	0.12	&	GLD3	\\
		MEC	&	1.75	&	3.78	&	0.67	&	0.22	&	0.06	&	SDI	\\
		NMC	&	1.11	&	13.40	&	0.32	&	0.08	&	0.09	&	GLD3	\\
		NVC	&	3.09	&	10.46	&	0.65	&	0.17	&	0.09	&	SDI	\\
		NXT	&	1.20	&	3.57	&	0.33	&	0.09	&	0.07	&	SDI	\\
		OMNI&	0.18	&	0.84	&	0.03	&	0.04	&	0.04	&	GLD0	\\
		PPC	&	0.88	&	7.07	&	0.26	&	0.09	&	0.08	&	SDI	\\
		XPM	&	0.99	&	0.86	&	0.20	&	0.05	&	0.05	&	SDI	\\
		QRK	&	1.15	&	1.74	&	0.36	&	0.08	&	0.10	&	GLD3	\\
		XRP	&	2.57	&	2.44	&	0.79	&	0.28	&	0.09	&	SDI	\\
		TAG	&	0.87	&	0.94	&	0.22	&	0.04	&	0.13	&	GLD3	\\
		TRC	&	0.71	&	0.96	&	0.09	&	0.04	&	0.02	&	SDI	\\
		WDC	&	1.80	&	5.03	&	0.70	&	0.23	&	0.10	&	SDI	\\
		ZET	&	0.92	&	2.33	&	0.25	&	0.05	&	0.09	&	GLD3	\\
		\bottomrule
	\end{tabularx}
	\caption{: Cram\'er von Mises Distance for different body model distributions.}
	\label{Tabelle_CMTest}
\end{table}
The results for the Anderson Darling distance are shown in Tab.\ \ref{Tabelle_ADTest} in the main text in Sec.\ \ref{sdiCC}.

\subsubsection*{Kolmogorov Smirnov distance measure}
Furthermore, we also determine the well-known distance between the empirical distribution function and the distribution model due to \citet{Kolmogorov.1933} and \citet{Smirnov.1936, Smirnov.1948}. The Kolmogorov Smirnov distance (KS) calculates the supremum of the absolute difference between the empirical and the estimated distribution functions.
Hence, the Kolmogorov Smirnov distance quantifies the "discrepancy" between the empirical distribution function of returns and the cumulative distribution function of the reference distribution, i.e.\ the model of the distribution function of CC returns
s. A more theoretical overview and comparisons to other distance measures can be found in, e.g.\, \citet{Stephens.1974, Shorack.2009}.

\subsubsection*{Kolmogorov Smirnov Distance}\label{Appendix_KSDistance}
\begin{table}[H]
	\footnotesize \noindent
	\centering
	\begin{tabularx}{0.75\textwidth}{l  r  r  r  r  r  r} \toprule
		CC								&	
		\multicolumn{5}{l}{Kolmogorov Smirnov Distances KS} &
		Best Choice							\\
		ID 									&	
		N									&	
		GED									&	
		GLD0  								&	
		GLD3							   	&	
		SDI									&
		{}									\\ 	\midrule
		EWCI$^-$	&	0.100	&	0.509	&	0.063	&	0.043	&	0.051	&	GLD3	\\
		ANC	&	0.132	&	0.233	&	0.069	&	0.070	&	0.047	&	SDI	\\
		BTB	&	0.087	&	0.156	&	0.056	&	0.037	&	0.042	&	GLD3	\\
		BTC	&	0.077	&	0.591	&	0.058	&	0.047	&	0.060	&	GLD3	\\
		CSC	&	0.179	&	0.296	&	0.085	&	0.053	&	0.047	&	SDI	\\
		DEM	&	0.071	&	0.127	&	0.048	&	0.027	&	0.039	&	GLD3	\\
		DMD	&	0.094	&	0.233	&	0.056	&	0.039	&	0.050	&	GLD3	\\
		DGC	&	0.117	&	0.241	&	0.066	&	0.034	&	0.045	&	GLD3	\\
		DOGE&	0.159	&	0.132	&	0.099	&	0.053	&	0.042	&	SDI	\\
		FTC	&	0.134	&	0.141	&	0.061	&	0.045	&	0.031	&	SDI	\\
		FLO	&	0.069	&	0.060	&	0.036	&	0.035	&	0.038	&	GLD3	\\
		FRC	&	0.166	&	0.250	&	0.097	&	0.061	&	0.047	&	SDI	\\
		GLC	&	0.072	&	0.190	&	0.038	&	0.025	&	0.040	&	GLD3	\\
		IFC	&	0.194	&	0.251	&	0.142	&	0.144	&	0.106	&	SDI	\\
		LTC	&	0.133	&	0.184	&	0.077	&	0.043	&	0.058	&	GLD3	\\
		MEC	&	0.144	&	0.206	&	0.096	&	0.070	&	0.041	&	SDI	\\
		NMC	&	0.095	&	0.368	&	0.056	&	0.040	&	0.037	&	SDI	\\
		NVC	&	0.167	&	0.338	&	0.092	&	0.052	&	0.042	&	SDI	\\
		NXT	&	0.134	&	0.196	&	0.077	&	0.051	&	0.039	&	SDI	\\
		OMNI&	0.052	&	0.113	&	0.028	&	0.031	&	0.036	&	GLD0	\\
		PPC	&	0.109	&	0.262	&	0.063	&	0.048	&	0.053	&	GLD3	\\
		XPM	&	0.116	&	0.095	&	0.055	&	0.048	&	0.035	&	SDI	\\
		QRK	&	0.108	&	0.139	&	0.073	&	0.048	&	0.048	&	SDI	\\
		XRP	&	0.171	&	0.168	&	0.101	&	0.061	&	0.039	&	SDI	\\
		TAG	&	0.101	&	0.103	&	0.056	&	0.037	&	0.048	&	GLD3	\\
		TRC	&	0.099	&	0.120	&	0.039	&	0.033	&	0.023	&	SDI	\\
		WDC	&	0.144	&	0.236	&	0.103	&	0.075	&	0.052	&	SDI	\\
		ZET	&	0.123	&	0.154	&	0.077	&	0.049	&	0.047	&	SDI	\\
		\bottomrule
	\end{tabularx}
	\caption{: Kolmogorov Smirnov Distance for different body model distributions.}
	\label{Tabelle_KSTest}
\end{table}

\section{Modelling the tail of a distribution}\label{Appendix_TailModel}
For a very large class of parent distribution functions, the GPD can be used as a model for the tail, cf.,\ e.g.,\ \cite{Embrechts.1997}. This class of distributions includes all common parent distributions that play a role in the financial sector. Because of that, almost no uncertainty exists regarding the model selection for the tail of the unknown parent distribution. The required quantiles can then be determined to high confidence levels with sufficient certainty. A certain threshold thus divides the parent distribution into two areas: a body and a tail region. This approach is already common practice for calculating high quantiles more accurately, as indicated in \citet{EuropeanParliament.2009}.

Various authors have proposed methods for determining the appropriate threshold and subsequently the GPD as model for the tail from empirical data. Most methods require the setting of parameters, which often requires experience and hinders full automation of the modeling process.
Based on ideas from \citet{Huisman.2001, Longin.2001, Chapelle.2005, Crama.2007, Hoffmann.2018, Hoffmann.2020} we use the following method for the determination of the tail model.

Given a suitable distance measure $\hat{R}_n = \hat{R}_n(F_n, \hat{F})$ as a function of the estimated GPD as a tail model $\hat{F}(x)$ and the empirical distribution function $F_n$ due to \citet{Kolmogorov.1933}, an automated modeling process can be constructed using the following algorithm.

\begin{enumerate}
	\item Sort the random sample taken from an unknown parent distribution in descending order: $x_{(1)} \geq x_{(2)} \geq \ldots \geq x_{(n)}$.
	\item \label{step2} Let $k =2,\ldots, n$, and find, for each $k$, the estimates of the parameters of the GPD. Note: For numerical reasons, we start at k = 2.
	\item \label{step3} Calculate the probabilities $\hat{F}(x_{(i)})$ for $i=1,\ldots,k$ with the estimated GPD, and determine the distance $\hat{R}_k$ for $k =2,\ldots, n$. 
	\item Find the index $k^*$ of the minimum of the distance $\hat{R}_k$.
\end{enumerate}
Then, the optimal threshold value is estimated by $\hat u = x_{(k^*)}$, and the model of the tail of the unknown parent distribution is given by the estimated GPD $\hat{F}(x)$, which itself is determined from the subset $x_{(1)} \geq x_{(2)} \geq \ldots \geq x_{(k^*)}$.

\citet{Chapelle.2005} suggest using the mean squared error (MSE) as distance measure $\hat{R}_n$. \citet{Bader.2018} implement a different algorithm based on sequences of goodness-of-fit tests. They use the Cram\'er-von Mises statistic and the Anderson-Darling statistic in combination with some beforehand set stopping rules.
In contrast, we use the statistics $AL_n^2$ (lower tail) or its counterpart $AU_n^2$ (upper tail) due to \citet{Ahmad.1988} as distance measure in the algorithm above. 

The aforementioned distance measures are basing on the the weighted mean square error Eq.\ (\ref{WMSEEQ})
\begin{comment} introduced in the context of statistical test procedures by \citet{Cramer.1928}, \citet{Mises.1931} and \citet{Smirnov.1936}. \end{comment} 
The distance measures due to \citet{Ahmad.1988} are derived from Eq.\ (\ref{WMSEEQ}) when the integral is calculated with the weight functions Eq.\ (\ref{WFEQ}) and $(a,b) = (1,0)$ for the lower tail (= $AL^2$) and $(a,b) = (0,1)$ for the upper tail (= $AU^2$). 

This choice is justified as follows:
The distance measure should take more account of the deviations between the measured data and the modeled data, especially in the tail area.
This can be achieved with the asymmetrical weight functions -- mentioned above -- used in the weighted mean square error as distance measure $\hat{R}_n$, cf.\ e.g.\ \citet{Hoffmann.2018, Hoffmann.2020}. So, the distance measures $AL_n^2$ and $AU_n^2$ of \citet{Ahmad.1988} take into account precisely the required asymmetrical weighting.

The pseudocode shown above was recently implemented in software packages and is available free of charge as a Python or R package, cf.\  {\color{red} Zitate, Links}. The analysis shown in the main part was carried out with this packages.

\newpage
%\section*{References}
%\bibliography{../../../FP_03_FatTailsCC/060_Publikationen/01_Tails/010_CryptoTailPaper_v01/LitCC_BundT}
\bibliography{LitCC_BundT}

\end{document}